\documentclass{aa}  

\usepackage{graphicx}
\usepackage{txfonts}
\usepackage{longtable}
\usepackage{natbib}
\bibpunct{(}{)}{;}{a}{}{,}
\usepackage[bookmarks=false, colorlinks=true, citecolor=blue, linkcolor=blue]{hyperref}
\def\kms{\,km\,s$^{-1}$}
\def\Jykms{\,Jy\,km\,s$^{-1}$}

\begin{document}

\title{Long-term multi-frequency maser observations of the intermediate-mass young stellar object G107.298+5.639 \thanks{Data used to produce Fig 1 and 2 are only available in electronic form at the CDS via anonymous ftp to cdsarc.u-strasbg.fr (130.79.128.5) or via http://cdsweb.u-strasbg.fr/cgi-bin/qcat?J/A+A/}}
\titlerunning{Maser observations in G107.298+5.639}
\authorrunning{M. Olech et al.}


   \author{M. Olech 
          \inst{1} \href{https://orcid.org/0000-0002-0324-7661}{\includegraphics[scale=0.5]{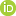}}
          \and
          M. Szymczak\inst{1} \href{https://orcid.org/0000-0002-1482-8189}{\includegraphics[scale=0.5]{orcid.png}}
          \and
                  P. Wolak \inst{1} \href{https://orcid.org/0000-0002-5413-2573}{\includegraphics[scale=0.5]{orcid.png}}
          \and
          E. G\'{e}rard \inst{2}
          \and
                  A. Bartkiewicz \inst{1} \href{https://orcid.org/0000-0002-6466-117X}{\includegraphics[scale=0.5]{orcid.png}}
}

   \institute{Institute of Astronomy, Faculty of Physics, Astronomy and Informatics, Nicolaus Copernicus University,\\
   Grudziadzka 5, 87-100 Torun, Poland 
         \and
         GEPI, UMR 8111, CNRS and Observatoire de Paris, 5 Place J.Janssen, 92195 Meudon Cedex, France\\
             }

   \date{Received month day , year; accepted month day, year}

\abstract
{Periodic flares of maser emission are thought to be induced either by variations of the seed photon flux in young binary systems or the pump rate regulated by stellar and accretion luminosities.} 
{We seek to study the variability of four maser transitions of three different species in G107.298+5.639 to constrain the dominant mechanism of periodic flares.} {Light curves of the 6.7\,GHz methanol and 22.2\,GHz water vapour maser were obtained with the Torun 32\,m radio telescope over 39 and 34 cycles, respectively. The target was also monitored at the 1.6\,GHz hydroxyl transitions with the Nan{\c{c}}ay radio telescope over 13 cycles. All these maser lines were imaged using VLBI arrays.}
{The study confirms alternating flares of the methanol and water masers with a period of 34.4\,d and reveals the synchronised behaviour of the methanol and hydroxyl masers in this source. The observed spatial distribution of the methanol maser cloudlets and the measured time delays of the flares of individual features imply a ring-like structure of radius 240\,au and thickness 30\,au. Internal proper motions indicate that the velocity of methanol cloudlets is dominated by a disc-wind component of about 5\kms. The methanol emission detected during only one VLBI observation is located in a region about 550\,au from a central star, which also exhibits OH maser flares. The erratic appearance of methanol features can be related to a powering object of relatively low luminosity which, during some variability cycles,  can excite molecules only in the nearest part of the disc. A careful  analysis of the maser and infrared light curves reveal a strong correlation between the 6.7\,GHz line and the infrared flux densities supporting a radiative pumping of the maser.}
{The synchronised behaviour of the hydroxyl 1665/1667\,MHz and 6.7\,GHz methanol transitions indicates a common pumping mechanism for the periodic flares of G107.298+5.639.}

   
   \maketitle
%
\section{Introduction}
Maser lines coming from outflows, circumstellar discs, and envelopes are the most characteristic properties of high-mass young stellar objects (HMYSOs) for a relatively very short ($5\times10^4$\,yr) phase of their evolution \citep{vanderwalt2005,Breen2010}. A small fraction of HMYSOs show periodic (24$-$668\,day) variations in the 6.7\,GHz CH$_3$OH maser emission \citep{Goedhart2003,Goedhart2004,Goedhart2009, Goedhart2014,Szymczak2011,Szymczak2015,Fujisawa2014,Maswanganye2015,Maswanganye2016,Sugiyama2017} 
and a few objects also show periodic variations in the H$_2$CO, OH and H$_2$O maser lines \citep{Araya2010,Al-Marzouk2012,Szymczak2016}.

Hypotheses explaining periodic variations in the methanol flux density are based on changes in the flux of seed photons or radiative pump rate. \cite{vanderWalt2011} claim that maser light curves of widely different shape and amplitude in some objects are due to periodic modulation of the background emission, which is amplified by the masering gas. Cyclic changes in the seed photon flux could arise in a colliding-wind binary or an eclipsing binary \citep{Van2009,vanderWalt2011,Maswanganye2016}. An alternative explanation of the  methanol periodicity is related to the pump rate, which depends on the flux of infrared emission \citep{Cragg2002}. \cite{Araya2010} propose a scenario of periodic heating of the dust due to gas accretion from the circumbinary disc onto protostars or accretion discs. Periodic maser flares could trace the accretion behaviour of circular or eccentric binaries \citep{Munoz2016}.  The dust temperature can be regulated by rotating spiral shocks in the gap of a circumbinary accretion disc \citep{Parfenov2014}. Flaring of the methanol masers in some objects can be also due to star pulsation driven by the $\kappa$ mechanism \citep{Inayoshi2013} or accretion instabilities developed by interactions between the stellar magnetosphere and the accretion disc \citep{D'Angelo2012}. 

Observations of periodic methanol masers at other maser transitions of well-established pumping schemes can shed more light on the nature of the periodicity. We can expect a correlation of periodic flares for the two maser transitions which share the same pumping mechanisms and anti-correlation if they are different. For the latter case a correlation should occur if the bursts are due to variations of the background emission amplified by the masers. \cite{Araya2010} report correlated quasi-periodic variability between H$_2$CO (4.8\,GHz) and CH$_3$OH (6.7\,GHz) masers in G37.55+0.20. In this object the flares of the excited OH (6.035\,GHz) masers were delayed relative to the H$_2$CO flares, but may be simultaneous with the CH$_3$OH peaks at corresponding radial velocities \citep{Al-Marzouk2012}. \cite{Green2012} report that the variability of the 1665\,MHz and 1667\,MHz lines in G12.889$+$0.489 shows evidence of correlation with the 6.7\,GHz line periodic variability. Furthermore, \cite{Goedhart2019} show that OH lines display concurrent flaring with 6.7 and 12\,GHz methanol masers in G9.62$+$0.20E. These tentative findings support a common excitation process of the OH and CH$_3$OH maser transitions. Recently, we found a remarkable pattern of anti-correlated alternating flares of the water (22\,GHz) and methanol (6.7\,GHz) maser transitions in the intermediate-mass young stellar object (IMYSO) G107.298+5.639 \citep{Szymczak2016}. These variations can be due to periodic changes in the infrared radiation field from the central protostar.

In this paper we report on our time-series observations of G107.298+5.639 (G107 hereafter) of maser transitions of three species to confirm the anti-correlated bursts of the methanol and water maser lines \citep{Szymczak2016} and to examine the behaviour of ground-state OH masers. The numerical models predict that the far-infrared pump together with the effects of line overlap provides the most convincing fit to the OH maser observational data (e.g. \citealt{Cesaroni1991,Gray1991,Cragg2002}). The OH and CH$_3$OH masers are radiatively pumped \citep{Cragg2002} and we report for the first time their correlated variability in this source. 

\section{Observations}
\subsection{Nan{\c{c}}ay and Torun radio telescopes}

The Nan{\c{c}}ay telescope has a half-power beam-width (HPBW) of 3\arcmin.5, in the E-W direction and 19\arcmin\, in the N-S direction. The beam efficiency is 0.65, the point source efficiency at zero declination 1.4\,K\,Jy$^{-1}$ and the system temperature is about 35 K.  All  four transitions of the OH ground state were observed simultaneously in both senses of circular polarisation using a 8192 channel auto-correlator spectrometer configured into eight banks of 1024 channels. A bandwidth of 0.39\,MHz was used for each bank, yielding a total useful velocity coverage of about $\pm$30\kms. The spectral bandwidth was centred at $-$15\kms\, relative to the local standard of rest (LSR). The spectra were taken in the frequency switching mode. The flux density calibration was based on the flux monitoring of W12 and is deemed to be accurate within 10\%. On four occasions around the anticipated maxima the 1665 and 1667\,MHz lines were observed in full polarisation mode following the procedure described in \cite{Szymczak2009}. The four Stokes parameters were obtained with a typical r.m.s noise level of about 60\,mJy in the $I$ Stokes and the uncertainty of the polarisation measurements was less than 8\%. The target was observed in the OH lines on average once a week and twice a month over 12 and 7 cycles, respectively.

The CH$_{3}$OH and H$_2$O maser lines were monitored with the Torun 32\,m radio telescope in the frequency-switching mode. For methanol the system temperature ranged from 25 K to 30\,K. The HPBW was 5\arcmin \, and the pointing accuracy was better than  $\sim$25\arcsec\, before mid-2016 and $\sim$10\arcsec\, afterwards. The data were calibrated by regular scans on the continuum source 3C123 and daily observations of the methanol maser G32.744$-$0.076, in which some features are non-variable within 5$\%$ \citep{Szymczak2014}. The typical r.m.s noise level in the final methanol spectra of 0.09\kms\, resolution was about 0.3\,Jy. The uncertainty of the flux density calibration was $\sim$10\%. The 6.7\,GHz emission was observed roughly every six hours and two to three days during the active and quiescent phases, respectively, over 39 cycles.

The 22\,GHz observations were calibrated by the chopper wheel method and observations of the continuum source DR21, assuming a flux density of 18.8\,Jy. The HPBW  was 1\farcm7 and  the system temperature ranged between 45 and 250 K depending on weather conditions and elevation. The typical r.m.s noise level was 15–20\,mK. The observations were corrected for atmospheric absorption and gain elevation effect. Estimated accuracy of temperature scale was about 20\%.  The water line was observed every one to two days over 33 cycles.

\subsection{Very Long Baseline Interferometry }
The dates of Very Long Baseline Interferometry (VLBI) experiments were adjusted according to the ephemeris from the 32\,m dish monitoring to get data just at flare peaks. The OH 1665 and 1667\,MHz maser lines were observed at two epochs: in 2017 November (MJD 58063) with the EVN (project ES083) and in 2017 June (MJD 57929) with the VLBA (project BS254A) at phases 0.56 and 0.63, respectively. The methanol maser line at 6.7\,MHz was observed  with the EVN (projects ES076, ES079A, ES079B, ES079C) at four epochs over the period from 2015 March to 2016 October at phases 0.50$\pm$0.03 (Table~\ref{vlbi-obs}). The H$_2$O line was observed  with the VLBA in 2017 June (project BS254A) at phase 0.97, i.e.  near the water flare peak. 

Table~\ref{vlbi-obs} presents an overview of all the observed maser transitions, including the arrays, dates, spectral resolutions, synthesised beam sizes, measured brightness of phase calibrator, and the r.m.s image noise. For all the experiments we used a phase-referencing observation scheme with J2223+6249 as a phase calibrator with a switching cycle of 195 s+105 s (maser + phase calibrator). 3C454.3 was used as a delay and bandpass calibrator. The typical total time on source was $\sim$5\,hr. Data were correlated with 1\,s integration time with the exception of experiment ES076, for which integration time was 0.25\,s.

The data reduction followed standard procedures for calibration of spectral line observations using the AIPS package. For EVN the initial step of  calibration was conducted on continuum pass and the solutions were applied to the spectral data sets, while VLBA calibrations were performed directly on spectral pass. The target was then self-calibrated on a channel representing the centre of the brightest spectral feature and the obtained phase and amplitude corrections were applied to the whole spectral data cube. For 1665\,MHz transition the emission in the reference velocity component was faint and spatially resolved out in long baselines (>1000\,km), so that the visibility calibration solutions were obtained from the data of the shorter baselines. Stations providing longer baselines or flagged owing to technical failures are denoted in italics in Table \ref{vlbi-obs}. No emission at 1667\,MHz was detected with both arrays.

Channel maps were searched for emission and the parameters of the detected maser emission were derived following the method presented in \cite{bartkiewicz2016} and \cite{Olech2019}. Point-like emission in each spectral channel is defined as spot. The position, intensity, and flux of each spot have been derived by fitting a 2D elliptical Gaussian. A group of spots coinciding within less than half of the synthesised beam size and present in adjacent channels was termed cloudlet (e.g.~\citealt{sanna2017}); their position was calculated as weighted average position of spots weighted by their measured flux. Cloudlets often formed smooth Gaussian spectral profiles and their parameters such as LSR peak velocity ($V_\mathrm{p}$), full width at half maximum (FWHM), peak brightness ($S_\mathrm{p}$), and fitted brightness ($S_\mathrm{f}$) were derived (Sect.~\ref{sec:mas-dis}). 
The absolute positional accuracy of the maser spots estimated using formula described in \cite{Bartkiewicz2009} are 18, 3, and 3\,mas
at 1.6, 6.7, and 22\,GHz, respectively.

To measure the proper motion of the methanol masers we used following procedure:
The data at all epochs were phase referenced and calibrated using AIPS package in an identical way. We used the first epoch as a reference to find the persistent maser cloudlets at the other epochs. Any systematic difference in the cloudlet position between the observations due to residual phase referencing errors, annual parallax, and source motion were removed by calculating the mean positions of cloudlet distributions, i.e. the barycentres of the masers. If after those correction any cloudlet drastically changed its position at any epoch, it was deemed as not persistent and removed from analysis. Finally the proper motion estimation of individual cloudlets was performed with linear fit of right ascension (RA) and DEC displacement over all epochs.

\begin{table*}
    \caption{Details on VLBI observations.}
    \label{vlbi-obs}
    \centering
    \begin{tabular}{l c c c c c c}
    \hline
     Telescope                      &  Date   & Frequency & Spectral res. & Synthesised                    &  $1\sigma_{\mathrm{rms}}$ & J2223+6249\\
                                    &  MJD    &  (MHz)    &    (\kms)     & beam ($\alpha\times\delta$; PA) &   per channel     &flux     \\
                                    &         &           &               & (mas$\times$mas;$\degr$)       &  (mJy\,beam$^{-1}$)& (mJy\,beam$^{-1}$) \\
    \hline
    {\bf EVN}{\tiny (Jb,Wb,Ef,Mc,O8,$T6$,$Ur$,$Tr$,$Sv$,$Zc$,$Bd$,$Ir$)} & 58063 & 1665.402   &     0.35              & 30$\times$23; 78.00            &   6    & 108               \\
                              &       & 1667.359   &     0.35              & 30$\times$23; 57.65            &   6                   \\
    {\bf VLBA}{ \tiny($BR$,FD,$HN$,KP,LA,$MK$,$NL$,$OV$,PT,$SC$)}   & 57929 & 1665.402   &     0.09              & 35$\times$33; 39.07            &   12        & 126           \\
                              &       & 1667.359   &     0.09              & 35$\times$33; 10.93            &   12                   \\   
    {\bf EVN}{\tiny (Jb,Ef,Mc,O8,Tr,$Nt$,Wb,Ys,$Sh$,$Sr$)} & 57097 & 6668.519 &  0.09              & 5.8$\times$3.7; -78.20            &   1.5      & 147             \\  
    {\bf EVN}{\tiny ($Jb$,Ef,Mc,O8,Tr,Nt,Wb,Ys,Sr)} & 57444 & 6668.519 &  0.09              & 5.2$\times$3.2; 61.38            &   3.0               & 189    \\ 
    {\bf EVN}{\tiny (Jb,Ef,Mc,O8,Tr,Nt,Wb,Ys,Sr,T6,Ir)} & 57544 & 6668.519 &  0.09              & 4.2$\times$2.8; 76.25            &  2.5  & 172  \\                          
    {\bf EVN}{\tiny (Jb,Ef,Mc,O8,Tr,Nt,Wb,Ys,T6)}       & 57681 & 6668.519 &  0.09             & 4.2$\times$3.1; 79.75            &  5.0   & 171 \\  
    {\bf VLBA}{\tiny(BR,FD,HN,KP,LA,$MK$,NL,OV,PT,SC)}     & 57906 & 22235.080&  0.11              & 0.6$\times$0.5; -1.10            &  5.0   & 104 \\   

    \hline
    \end{tabular}
    \tablefoot{No data were used from italicised stations}
\end{table*}

\section{Results}
\subsection{Maser light curves}
\begin{figure*}
   \centering
   \includegraphics[width=1.0\textwidth]{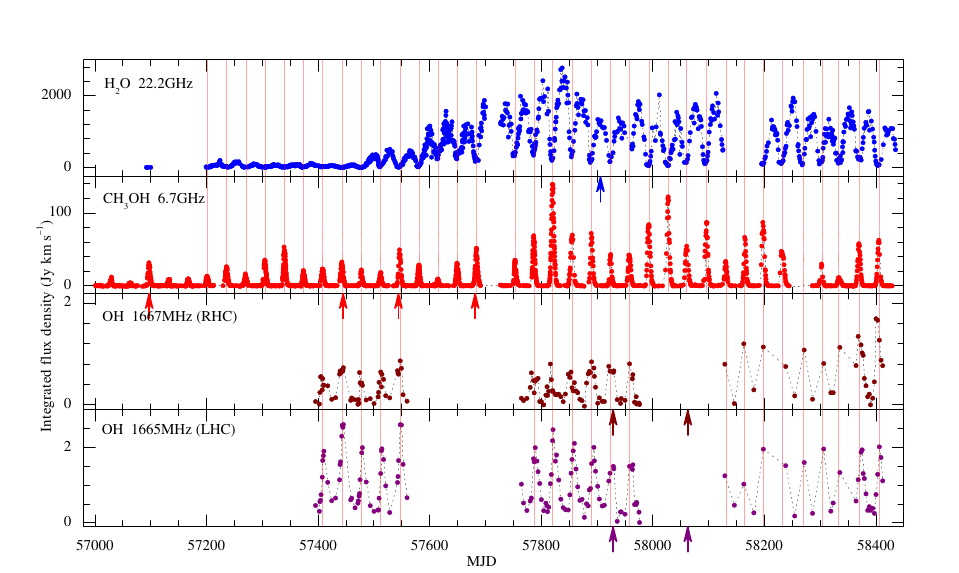}
   \caption{Time series of integrated flux densities in the $I$ Stokes for water vapour and methanol maser transitions and in the RHC and LHC  polarisation for hydroxyl maser transitions of G107. The vertical dotted lines indicate the methanol peaks. The arrows below the time axes denote the dates of VLBI observations (Table~\ref{vlbi-obs}).}
   \label{lc-all}
\end{figure*}

The integrated flux density time series for the four maser transitions presented in Fig.~\ref{lc-all} show clear periodic variations. The amplitude of individual features and shape of spectra displayed significant variability from cycle to cycle. These complex variations are best seen in the plots of the flux density as a function of time and velocity (Figs.~\ref{dyn-meth-wat} and \ref{dyn-water-OH}). 

Following the approach used in \cite{Olech2019} we calculated the period and analysed the flare profiles for all the transitions. An asymmetric power function of the form $S(t)=A^{s(t)}+C$ was fitted to the data. In this equation, $A$ and $C$ are constants and $s(t)=[b\mathrm{cos}(\omega t+\phi)/(1-f)\mathrm{sin}(\omega t+\phi))]+a$, $b$ is the amplitude relative to mean value $a$, $\omega=2\pi/P$, where $P$ is the period, $\phi$ is the phase, and $f$ is the asymmetry parameter \citep{Szymczak2011}. 
The resulting fits were used to assess the time of flare peak, the timescale of variability equals to the FWHM of the flare, the relative amplitude ($R$), and the ratio of the rise time to the decay time ($R_\mathrm{rd}$). Table \ref{Tab:line_properties} summarises their average values for all the spectral features. The active phase is the shortest for the 6.7\,GHz line; the average FWHM is only 4.30\,d and the flare profile is fairly symmetric with a  mean $R_\mathrm{rd}$=0.92. The flare profile at 22\,GHz has an average FWHM of 16.34\,d and the rise time is longer than the decay time ($R_\mathrm{rd}$=1.21). Both OH transitions have similar FWHM of $\sim$10\,d but the 1665\,MHz line has a steeper flare profile ($R_\mathrm{rd}$=0.61) than the 1667\,MHz line ($R_\mathrm{rd}$=0.93).

\begin{figure}[]
   \includegraphics[width=1\columnwidth]{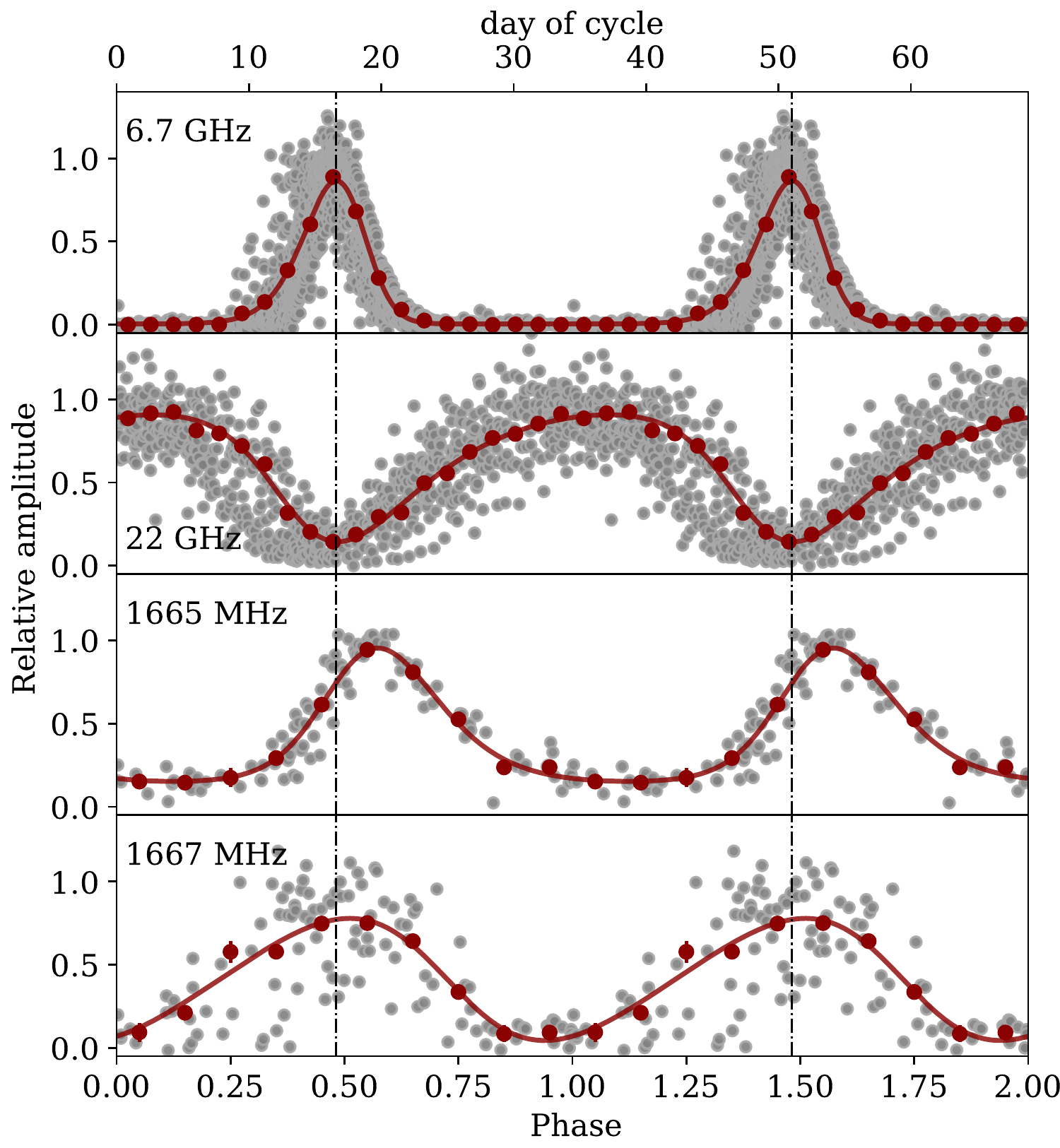}
   \caption{Phased and normalised light curves of the integrated flux  in each observed transition (grey points). Red points show binned light curves, with 0.05$P$ and 0.1$P$ bins for the 6.7, 22\,GHz and both OH lines, respectively. The red line represents the best fit of asymmetric periodic power function. All the light curves are folded modulo 34.43\,d, the variability period of the methanol emission, and the dashed vertical lines show the peaks of the 6.7\,GHz flare.}
   \label{phas_all}
\end{figure}

The time series of the velocity-integrated flux densities were normalised and phase-folded to show differences in the flare profiles and overall time delays between the peaks of observed transitions (Fig.~\ref{phas_all}). The anti-correlated behaviour of the water vapour and methanol masers is clearly visible; the minimum of water flux density coincides exactly with the maximum of methanol emission. 
Since the essential mechanisms of pumping H$_2$O and CH$_3$OH population inversions are collisional and radiative, respectively (\citealt{Elitzur1989, Cragg2002}), the observed anti-correlation rules out the hypothesis that the maser flares in the target are related to variations of seed photon flux caused by periodic changes in the free–free background emission in a system of colliding-wind binary. The methanol and OH 1665\,MHz light curves are quite a bit different, as the asymmetric model fits in Fig.~\ref{phas_all} suggest; the $R_\mathrm{rd}$ ratio is smaller at 1665\,MHz, where the peak is flat-topped followed by a slow decay as often seen in OH masers associated with the Mira variables (e.g. \citealt{Etoka2017}). The phased light curve of the 1667\,MHz emission is less reliable owing to a large scatter of the data  points and indicates no time lag relative to the methanol flare. 

\begin{table*}
  \caption{Characteristics of the maser flares of persistent features inferred from the long-term monitoring. The quantity $V$ is the peak velocity of the feature. Average values with their  standard errors are given for the flux density ($S$), relative amplitude ($R$), timescale of flare (FWHM), ratio of the rise time to the decay time ($R_\mathrm{rd}$), and time delay ($\tau$). 
  Delays for the methanol and water maser lines were calculated relative to the features $-$7.43 and $-$16.40\kms, respectively.
  }

    \label{Tab:line_properties}
    \centering
    \begin{tabular}{l c c c c c c}
    \hline
   Transition   & $V$ & $\overline{S}$ & $R$ &  FWHM & $R_\mathrm{rd}$ & $\tau$ \\  
      & (\kms) & (Jy) &  & (days) &  & (days)\\
     \hline
     CH$_3$OH 6.7\,GHz &$-$16.78 & 0.61 & > 6.00  & 2.99 (0.82) & 1.22 (0.39) & $-$2.09 (0.51)\\
     &$-$16.21 & 0.26 & > 2.10  & 3.10 (0.68) & 1.16 (0.32) & $-$1.94 (0.52)\\
     &$-$15.60 & 0.97 & > 7.34  & 5.03 (1.05) & 0.80 (0.23) & 3.67 (0.28)\\
     &$-$14.85 & 0.17 & > 0.94  & 5.03 (1.03) & 0.80 (0.20) & 1.43 (0.39)\\
     &$-$13.75 & 0.17 & > 1.38  & 4.08 (0.62) & 0.88 (0.16) & 2.95 (0.25)\\
     &$-$11.03 & 1.18 & > 4.70  &  5.82 (1.33) & 1.0 (0.43) & 1.58 (0.19)\\
     &$-$9.23 & 4.42 & > 31.29  &  3.78 (0.71) & 0.89 (0.24) & $-$0.11 (0.22)\\
     &$-$8.62 & 4.16 & > 19.77  &  4.73 (1.20) & 0.70 (0.26) & $-$1.93 (0.27)\\
     &$-$7.43 & 16.60 & > 105.81  &  4.10 (0.87) & 0.84 (0.22) & 0 \\
    \hline
   OH 1.665\,GHz &$-$12.22 & 1.65 & > 2.39  & 9.46 (0.60) & 0.67 (0.35) &  1.90 (0.53)\\
    &$-$10.03 & 0.44 & > 5.94  & 10.83 (1.06) & 0.54 (0.15) & 2.91 (0.44)\\
    \hline
  OH 1.667\,GHz  &$-$14.00 & 0.65 & > 1.41  & 9.87 (1.94) & 0.77 (0.38) & $-$2.63 (0.53)\\
    &$-$9.62  & 0.32 & > 1.25  & 11.10 (3.83) & 1.10 (0.50) & $-$1.02 (0.90)\\ 
     \hline
 H$_2$O 22\,GHz  &$-$18.83 & 42.00 & 22.49 (18.01) & 15.61 (2.32) & 1.29 (0.48) & $-$0.19 (0.41) \\
    &$-$16.40 & 218.60 & 146.51 (64.10) & 17.62 (1.49) & 1.19 (0.44) & 0 \\
    &$-$7.85 & 36.59 & 18.70 (26.68) & 15.79 (1.86) & 1.15 (0.49) & $-$0.50 (0.64) \\
    \hline
    \end{tabular}
   \end{table*}

\subsection{Periodicity estimates}
\begin{figure}[]
    
   \includegraphics[width=1\columnwidth]{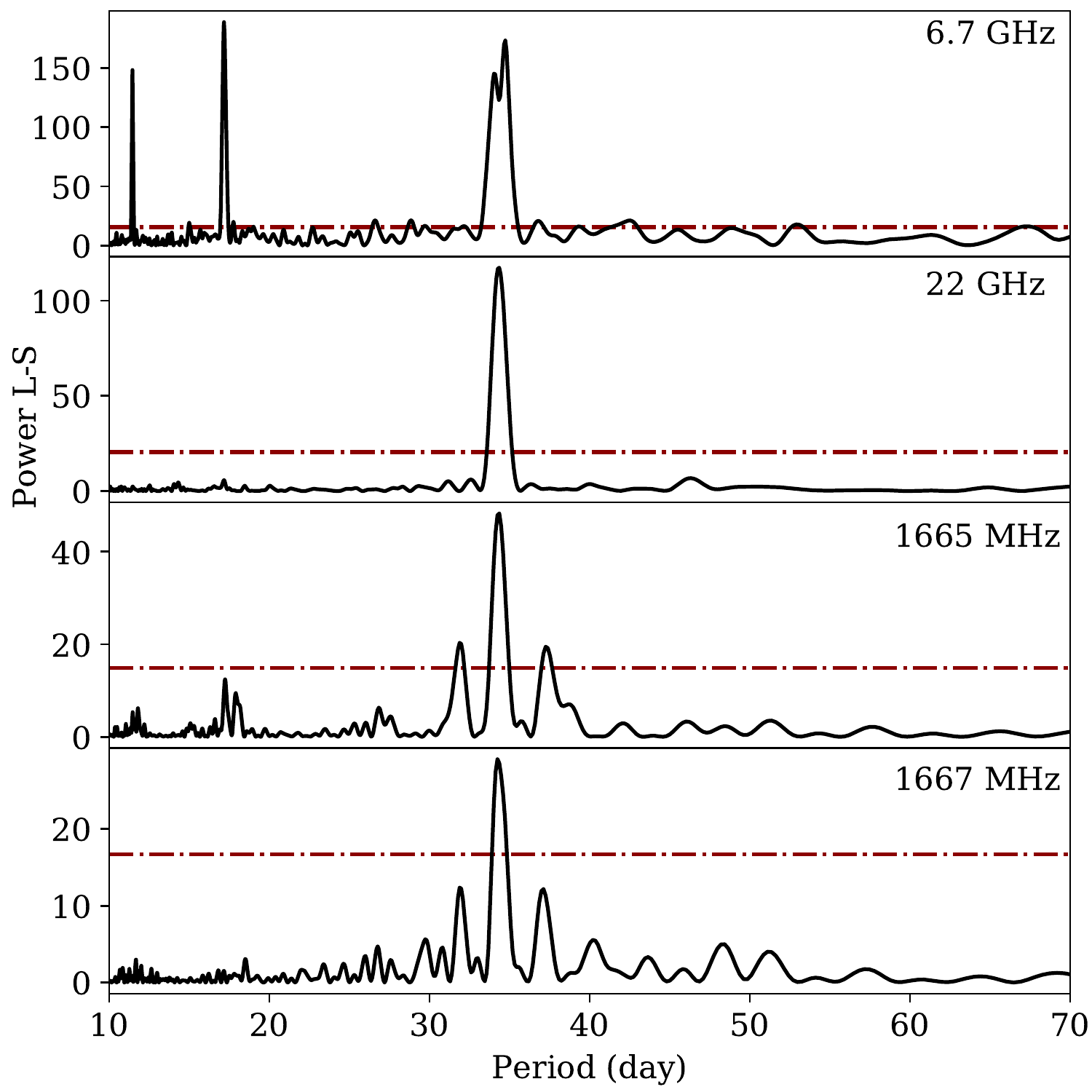}
   \caption{Lomb-Scargle periodograms (black lines) calculated for the integrated flux time series in each of the observed transitions. The red dashed line represents a 0.01$\%$ false alarm probability level.}
   \label{fig:perio}
\end{figure}

A Lomb–Scargle (L-S) periodogram as implemented in the VARTOOLS package (\citealt{vartools}) was used to search for any periodicity in the time series of the velocity-integrated flux densities in each of the four transitions shown in Fig.~\ref{lc-all}. We followed the procedure described in detail by \cite{Olech2019}. The L-S periodograms are shown in Fig.~\ref{fig:perio}. The periods with the highest power for the water and hydroxyl 1665 and 1667\,MHz masers are 34.35$\pm$0.51, 34.36$\pm$0.48 and 34.35$\pm$0.49 d, respectively. These values are the same, within the error range, as for the methanol maser of 34.41$\pm$0.72\,d.
The periodogram of the methanol time series clearly shows second and third harmonic signals of the derived period at
17.2 and 11.5\,d and the same responses are present for the 1665\,MHz signal as well. The peak corresponding to the methanol periodicity clearly splits as a consequence of slight variations in the flare profile parameters from cycle to cycle and varied contribution of individual features to the velocity-integrated flux density as Table~\ref{Tab:line_properties} depicts.

\subsection{Time delay}
We calculated the discrete correlation function (DCF; \citealt{Edelson1988}) to determine time delays between the flare peaks of persistent spectral features. The $-$7.43\kms\, feature of the 6.7\,GHz spectrum was chosen as a reference time series for each feature. The time delay was estimated by fitting a quadratic function to the DCF maximum as demonstrated in \cite{Olech2019}. The results are presented in the last column of  Table~\ref{Tab:line_properties} for the methanol and hydroxyl transitions. Thanks to high cadence observations and the relatively narrow profile of the methanol flare the estimated delays have an uncertainty of a fraction of a day. The time delays of the methanol features range from $-$2.1 to $+$3.7\,d, which well fit the line-of-sight effect for a typical size of maser source of about 1000 au (\citealt{Bartkiewicz2009}).
Time delays between the water maser features are less than 0.5\,d and suffer large uncertainties resulting from the strong blending effect of spectral components and widespread drifts in velocity.

\subsection{Polarised OH emission}
Figure~\ref{pol-OH1665} shows the full polarisation spectra of the OH 1665 and 1667\,MHz lines, respectively. Our convention for the calculation, scaling, and displaying of polarisation parameters is the same as in \citet{Szymczak2009}.

The 1665\,MHz emission ranges from $-$12.5 \kms\ to $-$7.4\kms\ with a peak flux density of 2.1\,Jy at $-$12.2\kms\, on MJD 57821, i.e. very close to the methanol maser maximum (Fig.~\ref{pol-OH1665}). Left-hand circular (LHC) polarisation emission dominates and the degree of circular polarisation ranges from 55\% to 95\%. The main feature at $-$12.2\kms\, is also linearly polarised up to 18\% and the polarisation angle is $-$75$\pm$10\degr. For all OH polarisation observations carried out very near to the methanol peaks at four consecutive cycles (from MJD 57787 to 57891) the polarisation angle remains constant within errors. The emission of the main 1665\,MHz feature was seen over the entire cycle of variability (Fig.~\ref{dyn-water-OH}). The 1667\,MHz spectrum is composed of two weak ($\la$0.5\,Jy) features near $-$14 and $-$9.5\kms. Only the right-hand circular (RHC) polarisation emission is detected. The 1667\,MHz flux density is above our detection limit of 240mJy only during the methanol flares. 

In general, the OH maser emission does not coincide in velocity with the methanol and water masers with the exception of a faint OH 1665\,MHz feature near $-$7.4\kms,\, which appears at the same velocity as the strongest methanol feature. The velocity-integrated flux density of the 1665\,MHz emission is a factor of 3 higher than that at 1667\,MHz.  
   
\subsection{Maser spatial distribution}\label{sec:mas-dis}

  \begin{figure*}[hbt]
   \sidecaption
   \includegraphics[scale=0.85]{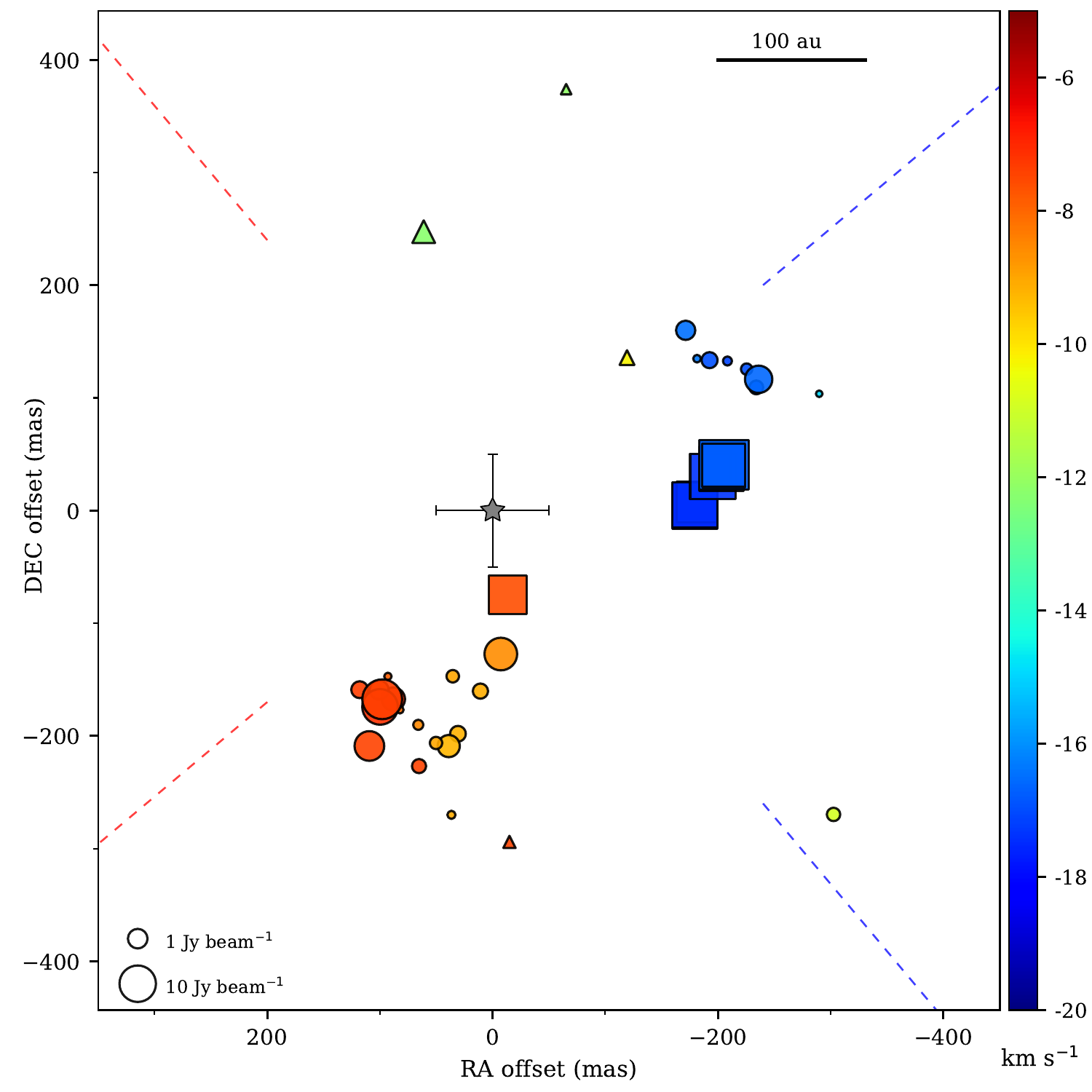}
   \caption{Composite map of three maser transitions in G107. The OH 1665\,MHz (triangles), CH$_3$OH 6.7\,GHz (circles), and H$_2$O 22\,GHz (squares) cloudlets are shown;    the observations were carried out on MJD 58063, 57681 and 57929, respectively.  The symbol size is proportional to the logarithm of maser brightness of  the cloudlets and its colour denotes the velocity according to the colour bar. The systematic errors in the relative position between different transitions are estimated to be less than 18\,mas. 
   The grey star symbol with the marked positional uncertainty represents the position of the 1.3\,mm continuum peak at RA(J2000)=$22^\mathrm{h}21^\mathrm{m}26\fs7730$ and DEC(J2000)=63\degr51\arcmin37\farcs657 (\citealt{Palau2013}) likely indicating the location of IMYSO. The blue and red dashed lines denote the directions of two large-scale  molecular outflows  (\citealt{Palau2011}). The horizontal bar indicates the linear scale for the assumed distance of 0.76\,kpc (\citealt{Hirota2008}).  \label{fig:vlbi_comp}}
   \end{figure*}

  Figure \ref{fig:vlbi_comp} shows the positions and radial velocities of the 22\,GHz water, the 6.7\,GHz methanol, and the 1665\,MHz hydroxyl masers observed very close to the maser peaks but at different cycles spanning less than 1.1\,yr. Table~\ref{tab:vlbi_char} presents the measured and derived maser properties: relative position, peak velocity, velocity FWHM of the cloudlet, brightness of the strongest spot of each cloudlet, and fitted peak brightness of each cloudlet.

During the  fourth epoch (MJD 57681) the methanol emission emerged from two groups of cloudlets $\sim$400\,mas apart along the NW$-$SE direction with position angle of $-$43\degr\, and a lonely cloudlet located in the SW part of the structure (Fig.~\ref{fig:vlbi_comp}). There is a clear segregation in radial velocity of the NW ($-$17.1 to $-$15.0\kms) and SE ($-$9.3 to $-$7.1\kms) groups of methanol masers that agrees well with the velocities of blue- and red-shifted lobes of outflows observed in the CO lines (\citealt{Palau2011}). 

The water maser emission was detected in two clusters placed within $\sim$200\,mas of each other just between the two main groups of methanol masers (Fig.~\ref{fig:vlbi_comp}). These water maser clusters differ in radial velocity in a similar way to those of methanol masers. It is interesting that the relative positions and radial velocities of these water maser clusters are similar to those seen ten\,years ago (\citealt{Hirota2008}). However, the morphology of the western cluster does not resemble that observed by \cite{Hirota2008}, who interpreted it as an expanding shell. Moreover, at the epoch of our observation (MJD 57901) no water maser was detected in the range from $-$3.6\kms to $-$3.1\kms and from $-$14.3\kms to $-$9.7\kms\, reported in \cite{Hirota2008}. 

The OH masers were successfully mapped only at 1665\,MHz. Three blue-shifted and one red-shifted cloudlet are detected at the north and south parts of the structure, respectively  (Fig.~\ref{fig:vlbi_comp}). Their distribution seen with different interferometers remains stable at two epochs spanning  4.4\,months. We note that the OH emission is completely resolved out at long baselines ($>$1000\,km) and the positions of cloudlets in Fig.~\ref{fig:vlbi_comp} denote the centroids of more extended structures of low brightness.

The overall structure of the red-shifted ($>-$11.3\kms) methanol maser remained the same at all four epochs. During the second epoch (MJD 57444) the blue-shifted emission 
had an entirely different structure than at the other three epochs (Fig.~\ref{comparison_methanol}); the emission from the NW location did not appear at all, but four new cloudlets at slightly different velocities emerged in the north about 500$-$600\,mas from the brightest red-shifted cloudlet near $-$7.4\kms. 
The ratio of the peak flares measured by the velocity-integrated flux density at the second and fourth epochs of VLBI observations was as 7:11. This ratio for the first and fourth VLBI epochs was 8:11, while the overall structure remained unchanged.
All four VLBI observations were taken less than 0.09 of phase around the flare maxima. This corresponds to 3 d and is a significant fraction ($\sim$30\%) of FWHM of feature flare. Thus the observed methanol morphology is highly dependent on the phase of the flare and the specific variability of the features for a given cycle as shown in Fig.~\ref{comparison_methanol-2}.

\subsection{Proper motion}\label{sec:pm}
In order to establish the correspondence of 6.7\,GHz maser cloudlets over the epochs, first we employed the two criteria proposed by \cite{Sanna2010}: (i) persistence of the relative distribution of a group of cloudlets and (ii) assumption of uniform motions. 
The second epoch data are excluded from this study since they did not meet the first criterion; the NW group of cloudlets was not seen while a new emission appeared about 600\,mas north of the persistent south structure. Thus for the purpose of proper motion measurements we used only the data from the rest three epochs. Following the procedure described in Sect.~2.2, we identified eight persistent cloudlets.  
 
Linear fits of displacements of the cloudlets over the epochs are shown in Fig.~\ref{fig:prop_jux} and the values of internal proper motions with fitting errors are listed in Table~\ref{tab:pm}. 

\begin{figure}
   \includegraphics[width=1\columnwidth]{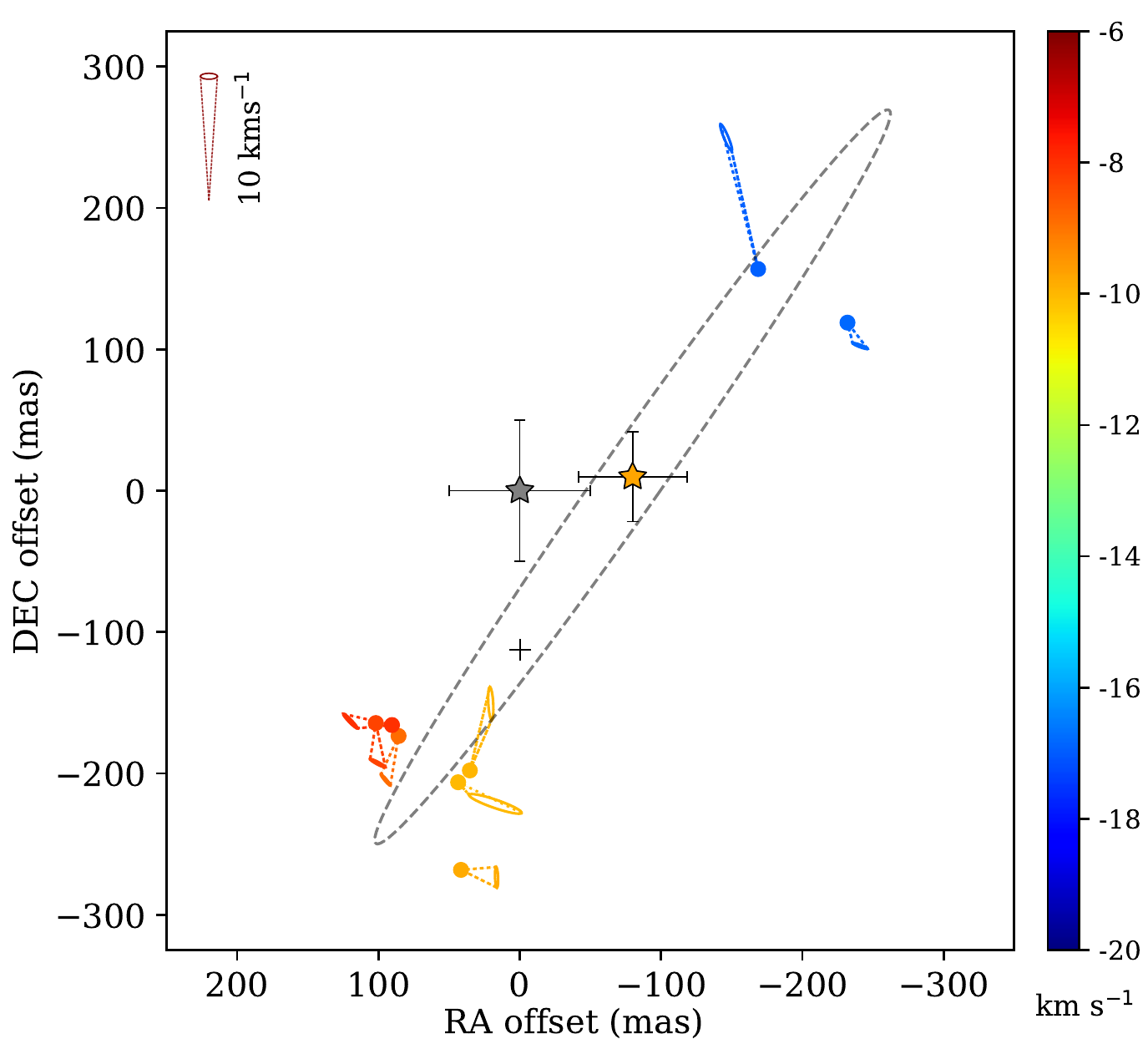}
   \caption{Proper motions of the methanol maser cloudlets in G107 as measured relative to their centre of
     motion denoted by small black cross. Cones represent the proper motions in each cloudlet listed in Table~\ref{tab:pm} and their colours and apertures correspond to the radial velocity (colour bar) and uncertainty, respectively. The velocity scale is shown at the top right corner. The dashed ellipse displays the model of ring derived
     from the 3D structure analysis shown in Fig.~\ref{vlbi_3d}, where the yellow star shows the position of fitted centre of the ring. The grey star denotes the position of the 1.3\,mm continuum peak (\citealt{Palau2011}).}
\label{fig:prop_mot}
\end{figure}

The measured internal proper motions are shown in Fig. \ref{fig:prop_mot}. The amplitude of proper motions ranges from 2.64$\pm$0.79\kms\, for cloudlet 7 to 13.41$\pm$1.43\kms\, for cloudlet 8. Both are blueshifted and lying at the NW. On average the cloudlets have a velocity of 5.45$\pm$3.21\kms. Most of the persistent cloudlets show motions nearly perpendicular to the plane of the edge-on disc-like structure, which best fits the 3D maser distribution and the time delays in the light curves of maser features as shown in Fig. \ref{vlbi_3d}. This possibly indicates that the methanol masers mainly trace the gas from a disc wind. However, the proper motions of the maser cloudlets in the SE part of the structure nearly align with the direction of the maser distribution (Fig.~\ref{fig:vlbi_comp}), indicating a flow along the structure, and may suggest a jet/outflow from the powering star.
Somewhat random proper motions in this site may result from the method of proper motion studies we applied. We calculated the proper motions with respect to the barycentre of methanol distribution rather than the exciting star.
We postpone a detailed analysis of plausible causes of the observed proper motions because the maser exhibits not only significant variations in the amplitude of flare and its profile, as the single dish monitoring proved, but also dramatic changes in the morphology at four cycles spanned only 1.6\,yr. We note that only about one-fourth of cloudlets was identified as persistent. It is also possible that variations in the relative intensity of nearby maser spots with similar radial velocities at the observed epochs mimic proper motion (\citealt{Sugiyama2016}). Further VLBI observations at high cadence could help interpret the  proper motions in the target.

\section{Discussion}
\subsection{Correlated variability of OH and CH$_3$OH masers} 
\begin{figure}
   \includegraphics[width=0.5\textwidth]{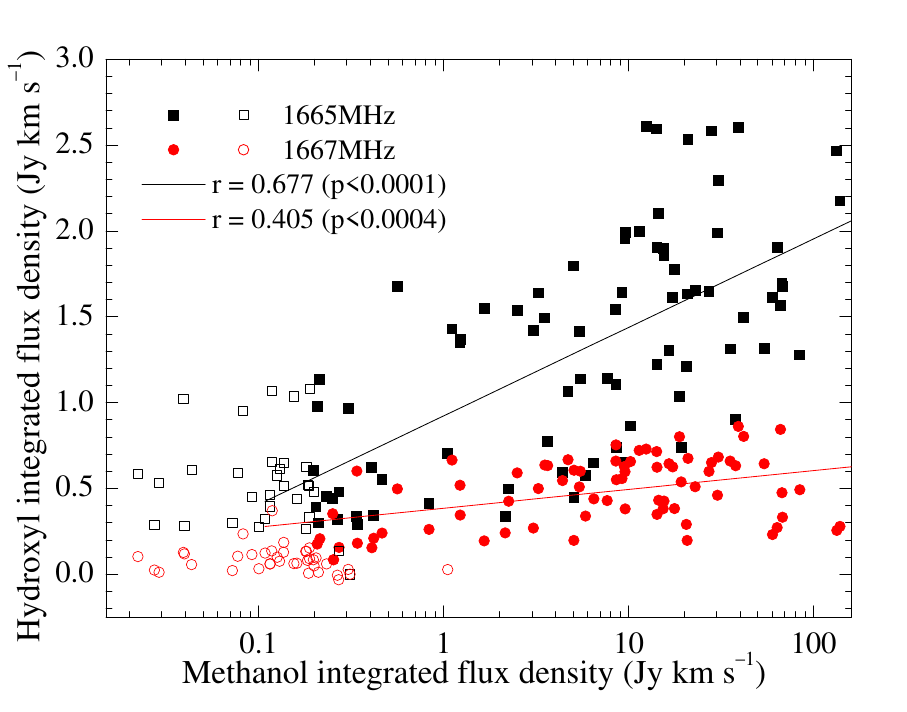}
   \caption{Correlation between the OH 1.665/1.667\,GHz and CH$_3$OH 6.7\,GHz maser integrated flux densities. The open symbols denote the upper limits for at least one transition. The solid lines show the best fits to the data.
   }
\label{cor-oh-ch3oh}
\end{figure}

Our observations definitively demonstrate that the intensity of the OH 1665 and 1667\,MHz masers tightly follows that of the CH$_3$OH masers (Fig.~\ref{lc-all}) with a $\la$0.1$P$ phase lag at 1665\,MHz (Fig.~\ref{phas_all}). For the time intervals of simultaneous observations of both species the integrated flux densities of the 1665/1667\,MHz and 6.7\,GHz lines are well correlated (Fig.~\ref{cor-oh-ch3oh}). This suggests similar excitation mechanisms for both molecules and that the masers come from the same gas volume.  

Fig.~\ref{fig:vlbi_comp} demonstrates that the OH maser cloudlet near $-$7.4\,\kms\,
lies less than 56\,mas (43\,au) from the CH$_3$OH cloudlet near $-$9.1\kms\, (Table~\ref{tab:vlbi_char}). The OH cloudlet near $-$10.3\kms\,
coincides within 57\,mas (44\,au) with the methanol cloudlet at $-$16.5\kms. In turn, the strongest and blue-shifted OH cloudlets at about $-$12.2\,\kms\, located $\sim$250\,mas north of the region of persistent methanol masers (Fig.~\ref{fig:vlbi_comp}) coincide within $<$120\,mas (100\,au) with an area at which the methanol maser emission of velocity between -15.6 and -13.7\kms\, appeared during the second epoch (MJD 57444) of VLBI observations (Fig.~\ref{comparison_methanol}). Since the flux densities of both masers are strongly affected by spatial filtering by the VLBI, as Fig.~\ref{comparison_methanol} depicts the above discussed spatial differences could be much smaller for low brightness extended emission. Thus, the hydroxyl and methanol masers originate in basically the same gas volume at a distance of 200$-$420\,au from the central protostar. We note that the projected size of $\sim$450\,au of the methanol maser region  during the three epochs is close to the  lower limit of the disc sizes observed in intermediate-mass embedded protostars \citep{Beltran2016}. In such environments the hydroxyl and methanol masers may probe gas of different kinematic properties excited by the common central object. Previous surveys revealed that the majority (70$-$80\%) of the 1.6\,GHz hydroxyl masers around HMYSOs have associated 6.7\,GHz methanol maser emission (\citealt{Caswell1998,Szymczak2004}) with many sites exhibiting a close spatial coincidence.

The mechanisms of OH pumping were widely discussed (e.g. \citealt{Guilloteau1981,Dewangan1987,Kylafis1990,Cesaroni1991,Cragg2002}) and there is now a consensus that far-infrared dust emission and line overlap are essential to a feasible pumping scheme, which accounts for the observed OH maser sources. In G107 we note the dominance of 1665\,MHz line over 1667\,MHz by a factor of three, which is typical for HMYSOs (\citealt{Caswell1987}) and implies the effect of line overlap (\citealt{Cragg2002}). The VLBI observations revealed that the ratio of brightness temperatures for the CH$_3$OH 6.7\,GHz and OH 1665\,MHz is 43 and 5.1 for the red- and blue-shifted cloudlets, respectively, implying a kinematic temperature of 110$-$120\,K and dust temperature $>$250\,K according to the model by \cite{Cragg2002} (their Figs. 5 and 6) with fixed values of the hydrogen density of $10^7$\,cm$^{-3}$ and OH column density of $10^{17}$\,cm$^{-2}$. Model calculations demonstrated that 6.7\,GHz methanol masers share a broad range of physical parameters with 1.6\,GHz hydroxyl masers (\citealt{Cragg2002}) and the coincidence of these transitions in G107 supports common excitation processes which produce the population inversions simultaneously in both molecules and result in synchronised variability of maser flux densities.

\subsection{Anti-correlated variability of CH$_3$OH and H$_2$O masers}
The present study reinforces our previous finding of alternating variations of the methanol and water masers in the source (\citealt{Szymczak2016}). The methanol maser flare exactly coincides with the water maser minimum for all 33 cycles monitored over $\sim$3.5\,yr (Figs.~\ref{lc-all}, \ref{phas_all}). Furthermore, the water maser emission that appeared outside the methanol velocity range, i.e. from $-$23 to $-$17\kms\, and greater than $-$5\kms, also shows periodic variability. All the water maser cloudlets lie well within the methanol bounded region of projected size less than 200\,au (Fig.~\ref{fig:vlbi_comp}). Thus, it is likely that the methanol and water masers operate in regions of similar physical parameters, where a factor causing periodic changes is efficient. 

The excitation mechanisms of the 6.7\,GHz and 22\,GHz masers are understood in general terms; the methanol line is likely inverted by infrared photons (e.g. \citealt{Sobolev1997,Cragg2002}) while the water line is predominantly collisionally pumped (e.g. \citealt{Elitzur1989,Gray2016}). Theoretical models generally indicate that the gas density and kinetic temperature favouring both maser transitions do not overlap (\citealt{Cragg2002,Gray2016}). However,  modelling of the 22 GHz maser excitation with up-to-date molecular data (\citealt{Nesterenok2013}) supports the view that the inversion is most efficient when the dust temperature is much lower than the gas temperature and an increase of the infrared emission reduces the maser gain (\citealt{Yates1997}). 

The methanol flare profile in G107 is tightly related to the infrared intensity variations (Sect.~\ref{s:met-IR}) supporting the radiative pumping models (\citealt{Cragg2002}). In response to a luminosity burst, traced by the methanol maser, the dust temperature reaches a maximum just at the methanol maser peak. This event reduces the ratio of the dust to kinetic temperature causing a decrease in the water maser intensity. As the heating and cooling times of dust grain are much shorter than those of gas (e.g. \citealt{Johnstone2013}) then we can expect the decaying branch of the H$_2$O light curve to be much steeper than the rising branch. This is exactly what we observed for the velocity-integrated emission (Fig.~\ref{phas_all}) and $R_\mathrm{rd}$ ratio of the maser features (Table~\ref{Tab:line_properties}).

\subsection{Three-dimensional model from VLBI maps and time delays}
   \begin{figure}
   \centering
   \includegraphics[width=1.0\columnwidth]{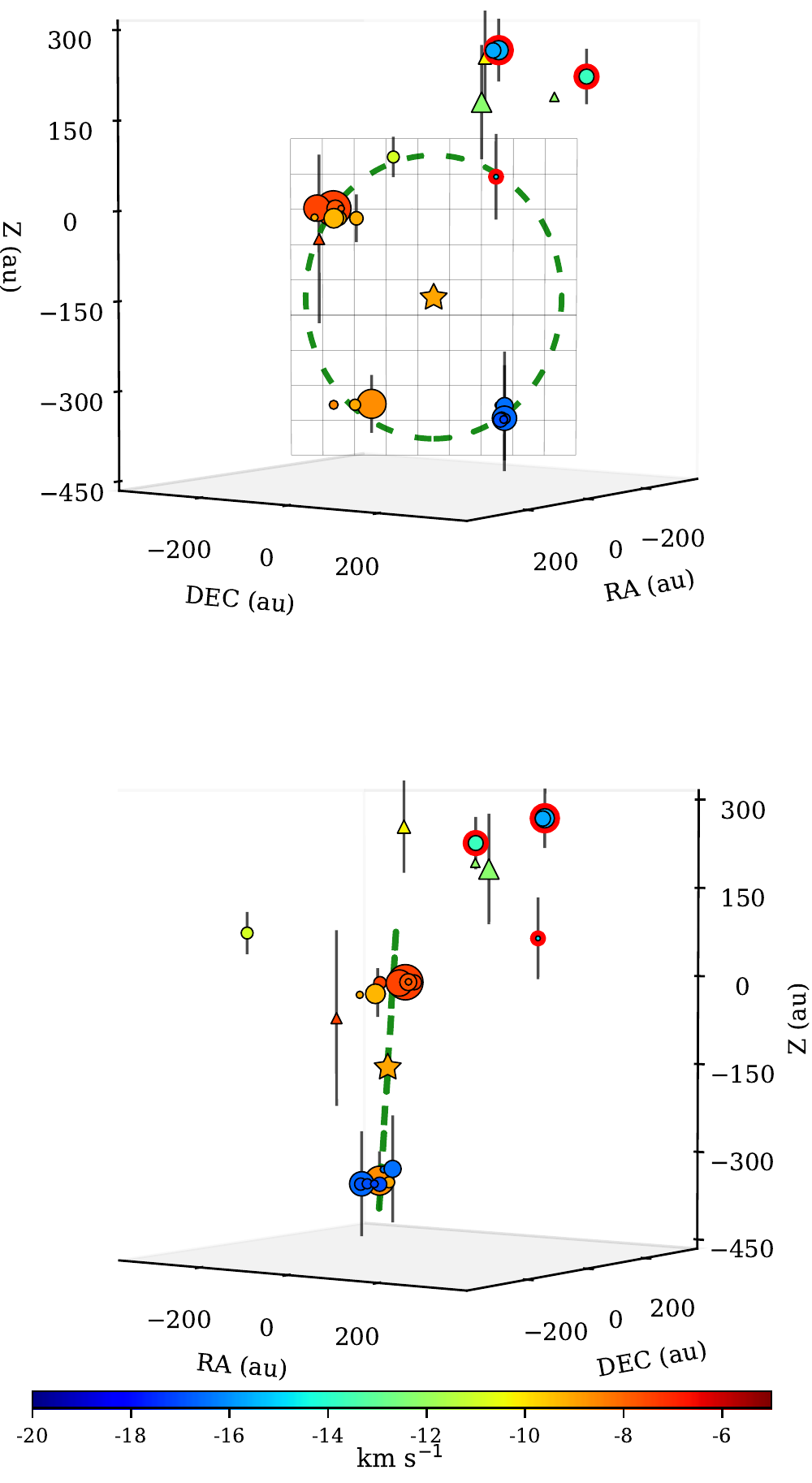}
   \caption{Three-dimensional maser structure of G107 obtained with the VLBI data and time delays of persistent feature peaks. The 6.7\,GHz methanol and 1665\,MHz hydroxyl cloudlets are shown as circles and triangles, respectively. The size of symbol is proportional to the logarithm of brightness and its colour corresponds to the LSR velocity scale shown in the horizontal wedge. The vertical bar for one cloudlet in each spectral feature denotes the uncertainty of the delay. The fitted position of protostar (orange star symbol) coincides within 1$\sigma$ with the measured position of the 1.3\,mm continuum emission (\citealt{Palau2011}). The top and bottom panels show  face-on and edge-on view of the ring-like structure.
   The green dashed circle in the top panel denotes the best fit to the methanol data at the fourth epoch of VLBI observation, whereas the green dashed line in the bottom panel indicates the plane of this ring-like structure.   The symbols surrounded by the red circles represent the methanol maser cloudlets seen only at the fourth epoch of VLBI observations. \label{vlbi_3d}}
    \end{figure}
In order to find the 3D methanol maser structure we applied the same procedure as in \cite{Olech2019}. The fourth epoch of VLBI data was used as the reference of the typical emission structure. The blue-shifted methanol emission seen only at the second VLBI epoch (Fig.~\ref{comparison_methanol}) and the OH emission mapped with VLBA were added to include all the methanol and hydroxyl features with measured time delays (Table~\ref{Tab:line_properties}). We assume that the periodic variations of the masers are caused by changes in the stellar and accretion luminosity of the central object which affect the infrared pump rate, while the time delays are due to differences in the photon paths between the cloudlets seen by the distant observer. We note that each feature in the spectrum was a superposition of the emission from spatially separated groups of cloudlets. Both the position of the central object and the line-of-sight distances between the cloudlets were fitted to the observed time delays. 

The recovered structure is presented in Fig \ref{vlbi_3d}. All persistent methanol features, with the exception of the $-$11.03\kms\, feature, lie in a plane inclined by 86$\pm$16\degr\, to the plane of the sky that is equidistant from the fitted position of powering object. The groups of cloudlets form a ring-line structure of radius 242$\pm$5.8\,au and an average thickness of 30.1$\pm$37.4au. This clearly indicates that the 6.7\,GHz maser emission comes from a disc seen almost edge-on. The $-$11.03\kms\, methanol emission lies at a distance of 360\,au from the protostar and above the fitted plane of the disc. The blue-shifted cloudlets detected at the second epoch of VLBI observations are 500$-$560\,au from the  central protostar. The erratic appearance of these cloudlets in VLBI observations and low relative amplitudes of variability seen in the single-dish monitoring are possibly related to a relatively low luminosity of the central object. We can expect that flares of the central star at only some cycles could be strong enough to heat up the dust in those distant regions of the circumstellar envelope.
In this model the red-shifted OH emission ($-$7.4\kms) lies close to the methanol red-shifted cloudlets, while that at blue-shifted  velocities coexists with the erratic
(seen during the second VLBI observation) methanol cloudlets at about 450\,au from the central star.
We conclude that assuming  radiative pumping of the CH$_3$OH and OH masers,  the structure of maser emissions and time delays of the flares of individual features observed in G107 can be satisfactorily explained by the disc model.

\subsection{Relationship between the methanol maser and IR variability}\label{s:met-IR}
To explore the  relationship between the maser flares and infrared emission in G107 we extracted the 3.4\,$\mu$m (W1) and 4.6\,$\mu$m (W2) time series from NEOWISE photometry.  Details about the instrument and observations are given in \cite{Mainzer2011,Mainzer2014} and a guide for users of the NEOWISE data is presented by \cite{Cutri2015}. The NEOWISE-R Single Exposure Source Table\footnote{Available at irsa.ipac.caltech.edu} was queried using a $5\arcsec$ search radius and the data covering the maser monitoring period were found for eight observation cycles  1.3-2.0\,d in length (hereafter called epochs). During each epoch, some 20-30 points were available with intervals of 0.066 to 0.262 d. The data with the frame quality score {\it qual$_-$frame}= 0 were rejected and 211 exposures of high quality ({\it ph$_-$qual} = "A") were used. Their median signal-to-noise ratios at bands W1 and W2 are 47.6 and 83.4, respectively. The magnitudes of individual measurements were converted to fluxes following the WISE Explanatory Supplement\footnote{wise2.ipac.caltech.edu/docs/release/neowise/expsup/}.

\begin{figure*}
   \includegraphics[width=1.0\textwidth]{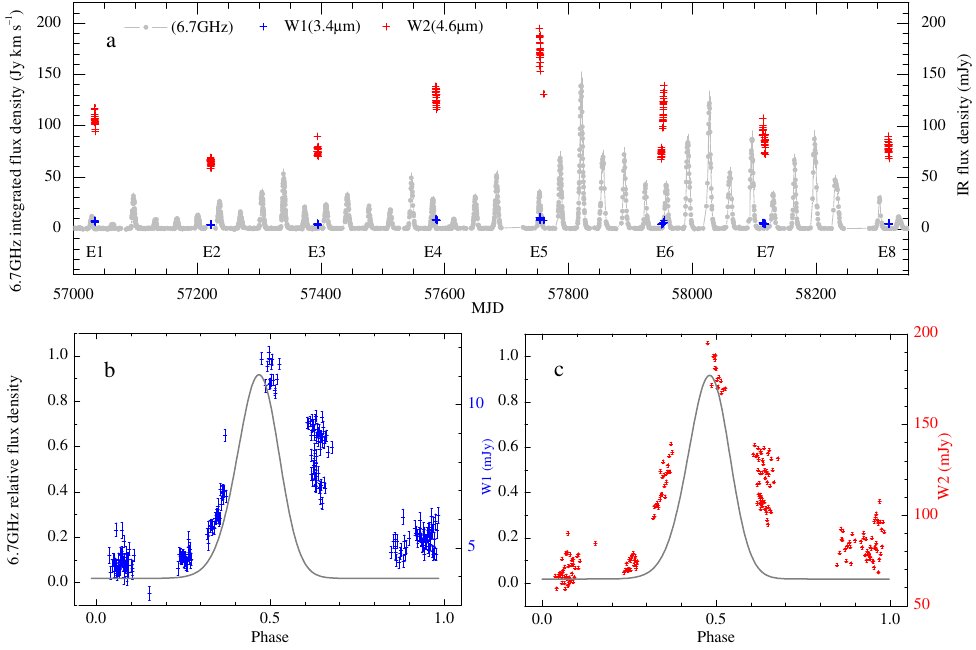}
   \caption{{\bf a.} Comparison of the methanol maser time series and the NEOWISE photometric data during this period. The epochs of IR observations are numbered with E1 to E8. {\bf b.} and {\bf c.} NEOWISE light curves folded by the 34.4 d period for bands W1 and W2, respectively. The grey line shows the phased and normalised methanol light curve as taken from the top panel of Fig.~\ref{phas_all}.
   }
\label{m-ir-lc}
\end{figure*}

Fig.~\ref{m-ir-lc}a shows the time series of the 6.7\,GHz integrated flux density and 3.4 and 4.6$\mu$m flux densities. It is clear that the NEOWISE observations at the second (E2, $\sim$MJD 57222), third (E3, $\sim$MJD 57395), seventh (E7, $\sim$MJD 58116), and eighth (E8, $\sim$MJD 58317) epochs were taken during a quiescent state, when the maser flux density was below our detection limit. For the remaining four epochs the IR observations were contemporaneous with active phases of the maser. We note that the amplitude of the maser flares closest to the epochs of NEOWISE observations ranged from 10 to 40\,\Jykms, while over the whole monitoring period (Fig.~\ref{lc-all}) it varied from 3 to 150\,\Jykms.  This may suggest that a long-term variability does not strongly affect the maser$-$infrared relationship at these eight epochs. 

\begin{figure}
   \includegraphics[width=0.5\textwidth]{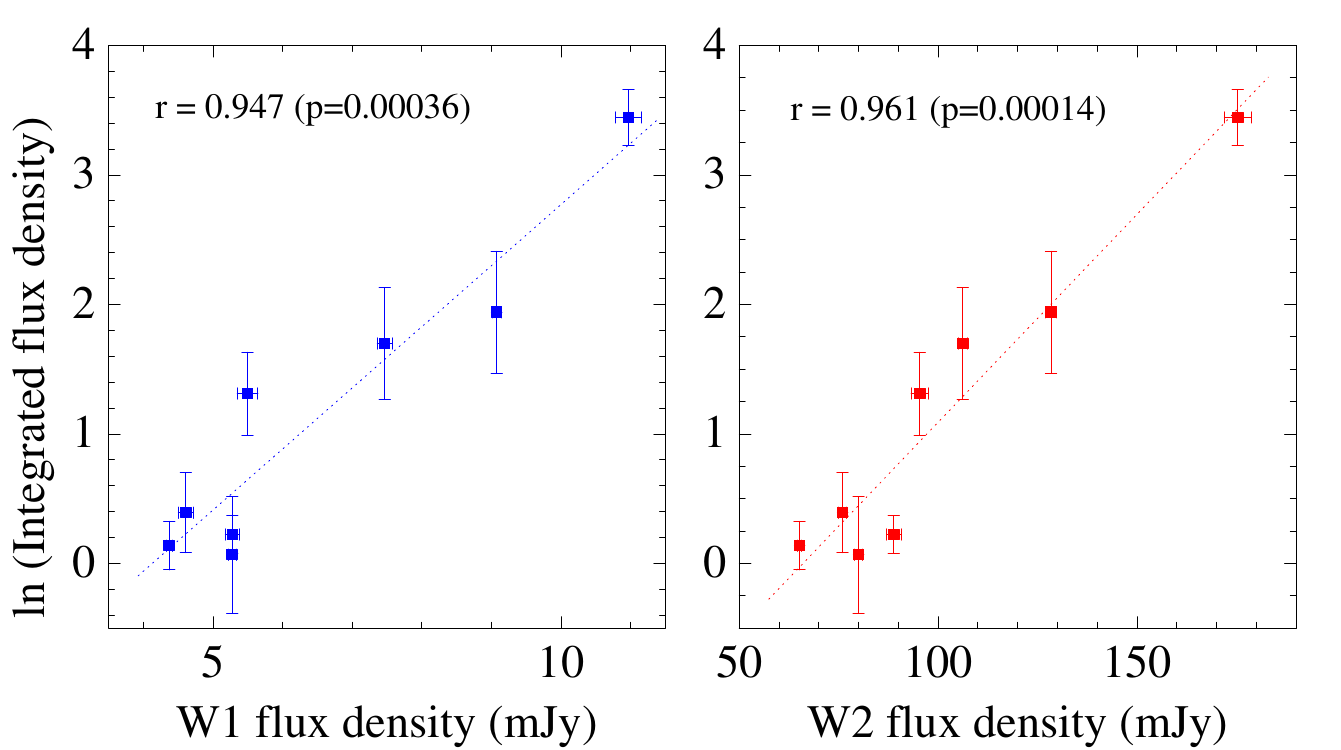}
   \caption{Natural logarithm of the methanol integrated flux density measured in \Jykms\, vs. band W1 (blue) and W2 (red) flux densities. The dotted lines denote least-squares fits to the data. 
   }
\label{m-wise}
\end{figure}

The phased infrared light curves are shown in Figs.~\ref{m-ir-lc} b and c. There is a general agreement between the flare profiles at the infrared and maser wavelengths 
with the exception of a bump in the IR light curves at phase $\sim$0.63. The NEOWISE data at this phase come from epochs E1 and E4  (Fig.~\ref{lc-all}a), when the maser peaks during the preceding activity cycles differed by a factor of 2.9. The ratio of average flux densities at E4 and E1 epochs is 1.21 for both IR bands. If we assume
that this change in the IR fluxes makes a difference in the maser optical depth $\Delta\tau$=1.21 then for unsaturated maser regime we obtain e$^{(\Delta\tau)}$=3.3, which is surprisingly consistent with the observed ratio of the maser peak flux densities of 2.9. This appears consistent with a scenario of pumping by infrared emission from dust surrounding the  central star \citep{Sobolev1997,Cragg2002}.
\cite{Stecklum2018} have established the light curve of G107 from ALLWISE and NEOWISE data folded by a 34.6\,d maser period. The saw tooth shape of their curve differs from that presented here, since they did not have precise information about the phase of the maser flares. Fig.~\ref{m-wise} confirms the  strong correlation between the 6.7\,GHz line and IR flux densities and supports a radiative pumping of this unsaturated maser emission. Our results could  be further refined using high cadence data at longer IR wavelengths to investigate in more details the pumping mechanisms.

\section{Summary and conclusions}
The IMYSO\ G107 was observed at high angular resolution and monitored over the period 2015-2018 in the four maser transitions: 22.2\,GHz H$_2$O, 6.7\,GHz CH$_3$OH, 1665 and 1667\,MHz OH. The main findings and conclusions are summarised as follows:
\begin{enumerate}[1.]
    \item The periodic variations of the OH and CH$_3$OH masers are simultaneous to within less that 3\,d, whereas those of the H$_2$O maser are delayed exactly by half of the period of 34.4\,d, i.e. the water maser emission is significantly dimmed or even disappears at regular intervals just when the OH and CH$_3$OH masers peak.  
    \item We found a strong correlation between the 6.7\,GHz integrated flux density and 3.4 and 4.5\,$\mu$m flux densities observed with the NEOWISE at eight epochs. This strongly supports theoretical models of radiative pumping of this maser transition.
    \item VLBI observations confirmed the elongated distribution of the methanol emission with a clear velocity gradient. The 3D structure recovered from these data and time delay measurements of persistent features implies a ring-like structure of radius 240\,au and thickness 30\,au, while erratic methanol emission seen at one epoch is located
   500\,au from the exciting star. Thus the observations are well explained with a disc model in which both OH and CH$_3$OH masers are radiatively pumped.   
    \item Measurements of internal proper motions of the methanol maser cloudlets indicate velocity vectors mostly orientated perpendicularly to the main axis of structure suggesting the presence of a disc wind.
\end{enumerate}

Overall, the study provides strong support for models of radiative pumping of the OH and CH$_3$OH molecules in a working region between the disc and outflow. Future VLBI imaging will help constrain the kinematics of cloudlets of all the maser transitions and quasi-simultaneous observations in other radio and infrared bands, and allow us to better understand the complicated structure seen in the thermal lines (e.g. \citealt{Palau2013}).

\begin{acknowledgements}
This work is based in part on observations carried out using the 32-meter radio telescope operated by Torun Institute of Astronomy, Nicolaus Copernicus University in Torun (Poland) and supported by the Polish Ministry of Science and Higher Education SpUB grant. 
This work was supported by the National Science Centre, Poland through grant 2016/21/B/ST9/01455. The Nan\c{c}ay Radio Observatory is operated by the Unit\'e Scientifique de Nan\c{c}ay of the Observatoire de Paris, associated with the CNRS. The European VLBI Network is a joint facility of independent European, African, Asian, and North American radio astronomy institutes. The VLBA is an instrument of the Long Baseline Observatory, which is a facility of the National Science Foundation operated by Associated Universities, Inc. This publication also makes use of data products from NEOWISE, which is a project of the Jet Propulsion Laboratory/California Institute of Technology, funded by the Planetary Science Division of the National Aeronautics and Space Administration.

\end{acknowledgements}

\bibliographystyle{aa}
\bibliography{librarian-g107_v2.bib}

\begin{appendix}

\section{Supplementary tables and figures}

\setcounter{table}{1}

\begin{table*}
\caption{Proper motions of the 6.7\,GHz maser cloudlets in G107. \label{tab:pm}}
\centering
\begin{tabular}{c r r c c c c c c}
    \hline
Cloudlet&$\Delta$RA & $\Delta$DEC & $V_\mathrm{p}$  & $S_\mathrm{p}$ & $S_\mathrm{f}$ & V$_\mathrm{RA}$ & V$_\mathrm{DEC}$  \\
     &(mas) & (mas) & (\kms)  & (Jy~beam$^{-1}$)  & (Jy~beam$^{-1}$) & (\kms) & (\kms) \\
    \hline
1 & 90.409 & $-$165.812 & $-$7.12  & 0.119 & 0.121 & 4.14 $\pm$ 0.70 & 0.40 $\pm$ 0.72\\
2 & 101.875 & $-$164.401 & $-$7.42  & 7.575 & 8.181 & $-$0.19 $\pm$ 0.76 & $-$3.98 $\pm$ 0.40\\
3 & 85.773 & $-$173.574 & $-$8.09  & 0.053 & 0.056 & 1.28 $\pm$ 0.46 & $-$4.36 $\pm$ 0.55\\
4 & 41.593 & $-$268.280 & $-$9.07  & 0.048 & 0.052 & $-$3.53 $\pm$ 0.04 & $-$0.74 $\pm$ 1.02\\
5 & 35.316 & $-$197.933 & $-$9.25  & 0.187 & 0.196 & $-$2.07 $\pm$ 0.09 & 6.64 $\pm$ 1.62\\
6 & 43.614 & $-$206.295 & $-$9.28  & 0.328 & 0.321 & $-$3.69 $\pm$ 2.60 & $-$2.14 $\pm$ 0.89\\
7 & $-$231.936 & 118.904 & $-$16.58  & 0.016 & $-$ & $-$1.26 $\pm$ 0.74 & $-$2.32 $\pm$ 0.28\\
8 & $-$168.709 & 156.797 & $-$16.68  & 0.032 & 0.033 & 3.17 $\pm$ 0.55 & 13.03 $\pm$ 1.32\\
\hline
\end{tabular}
\tablefoot{The values of internal proper motions of the cloudlets and their uncertainties are listed in columns 7 and 8. Designations of columns 2$-$6 are identical to those in Table~\ref{tab:vlbi_char}.}
\end{table*}

\begin{figure*}
   \centering
   \includegraphics[width=0.92\textwidth]{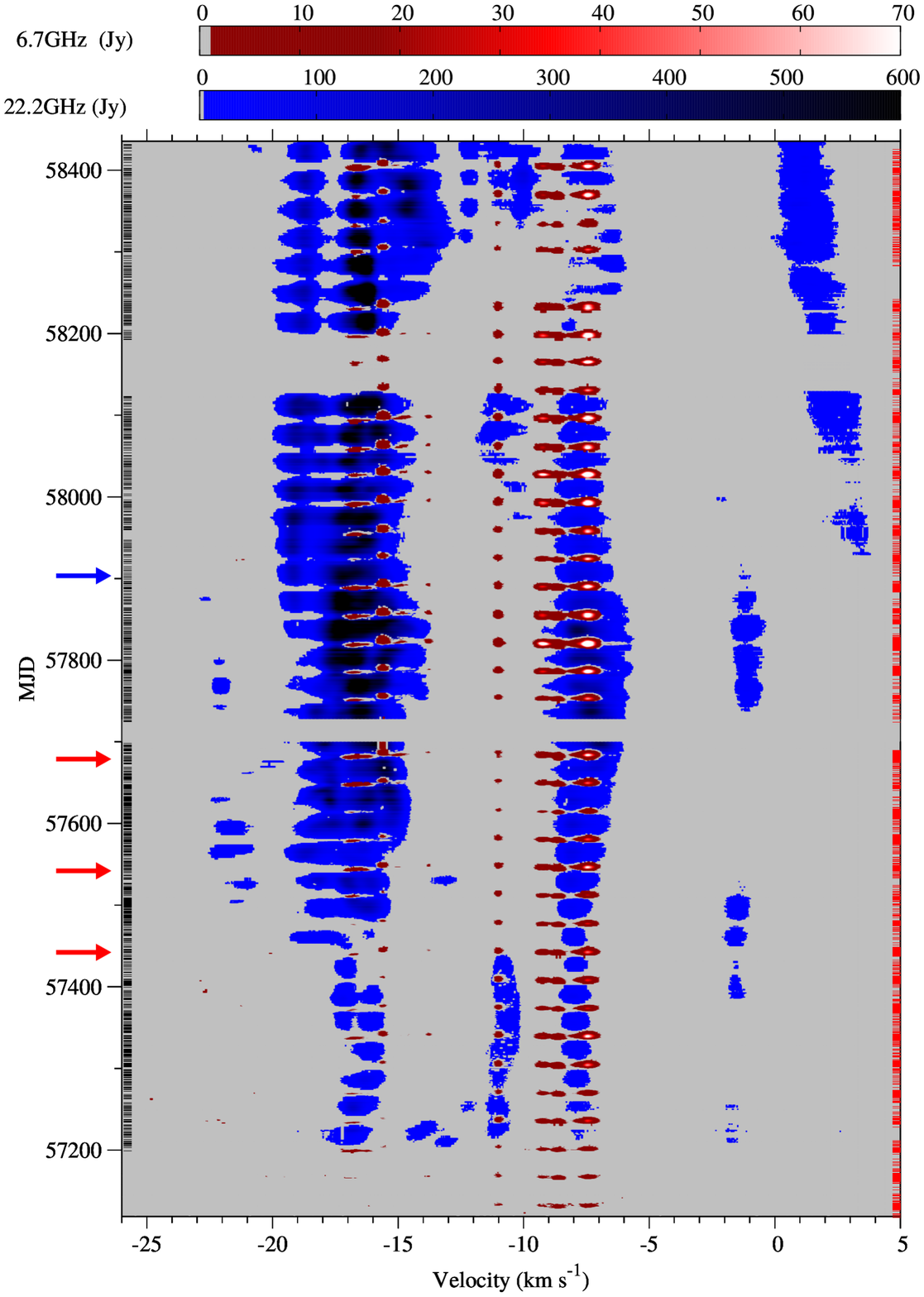}
   \caption{Dynamic spectra of the methanol and water maser lines of G107. Radial velocity is measured with respect to the LSR. The horizontal bars in the left (black) and right (red) ordinates correspond to the dates of the observed spectra of water and methanol lines, respectively.
    The red and blue arrows indicate the epochs of VLBI observations of methanol and water, respectively. 
   }
   \label{dyn-meth-wat}
\end{figure*}
 
\begin{figure*}
   \centering
   \includegraphics[width=0.92\textwidth]{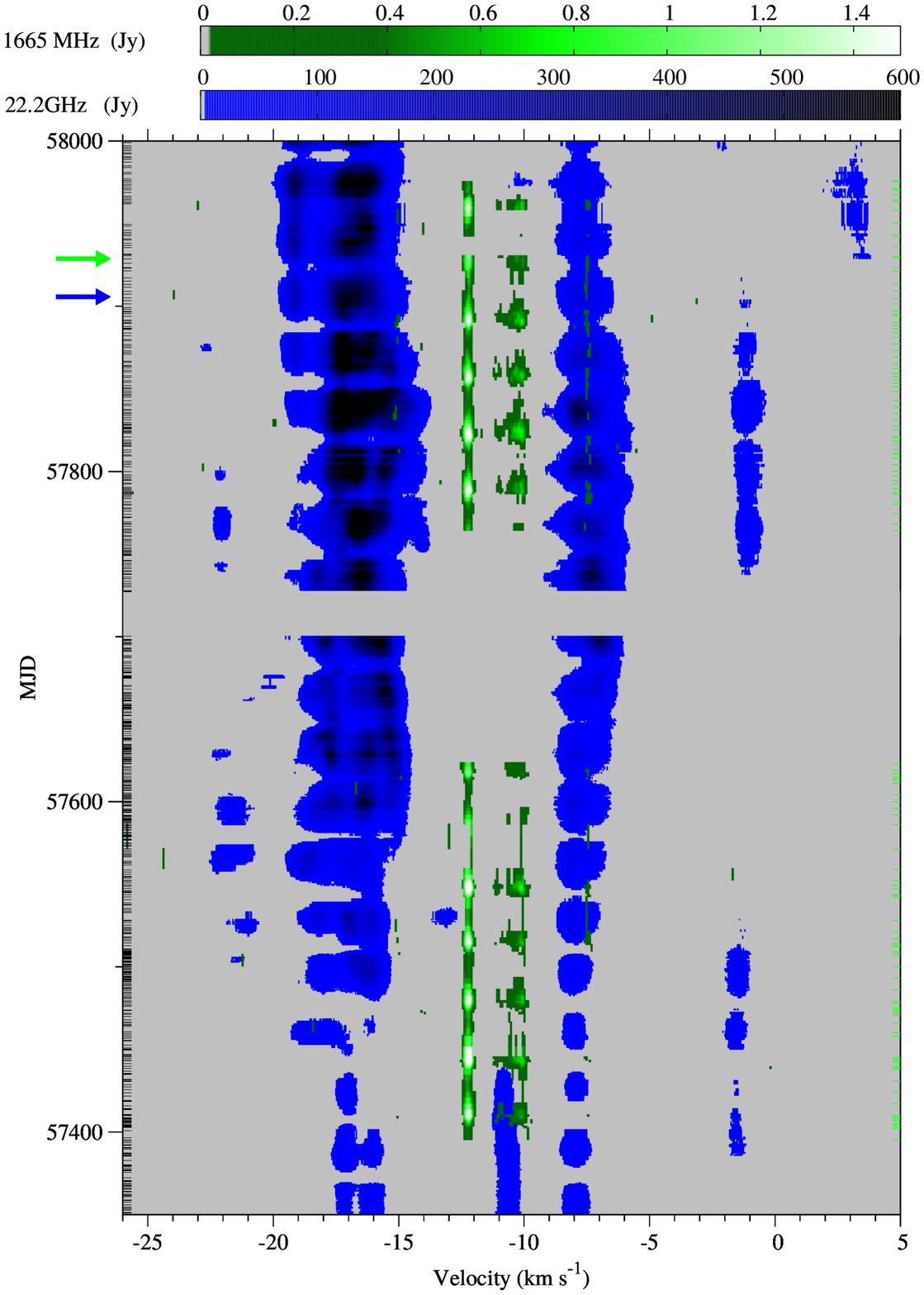}
   \caption{Dynamic spectra of the water and hydroxyl maser lines of G107. Radial velocity is measured with  respect to  the LSR. The horizontal bars in the left (black) and right (green) ordinates correspond to the dates of the observed spectra of water and hydroxyl lines, respectively.
   The blue and green arrows indicate the epochs of VLBI observations of water and hydroxyl, respectively.
   }
   \label{dyn-water-OH}
\end{figure*}

\begin{figure}
   \includegraphics[scale=0.5, trim={1.8cm 1.20cm 3.2cm 2.0cm},clip]{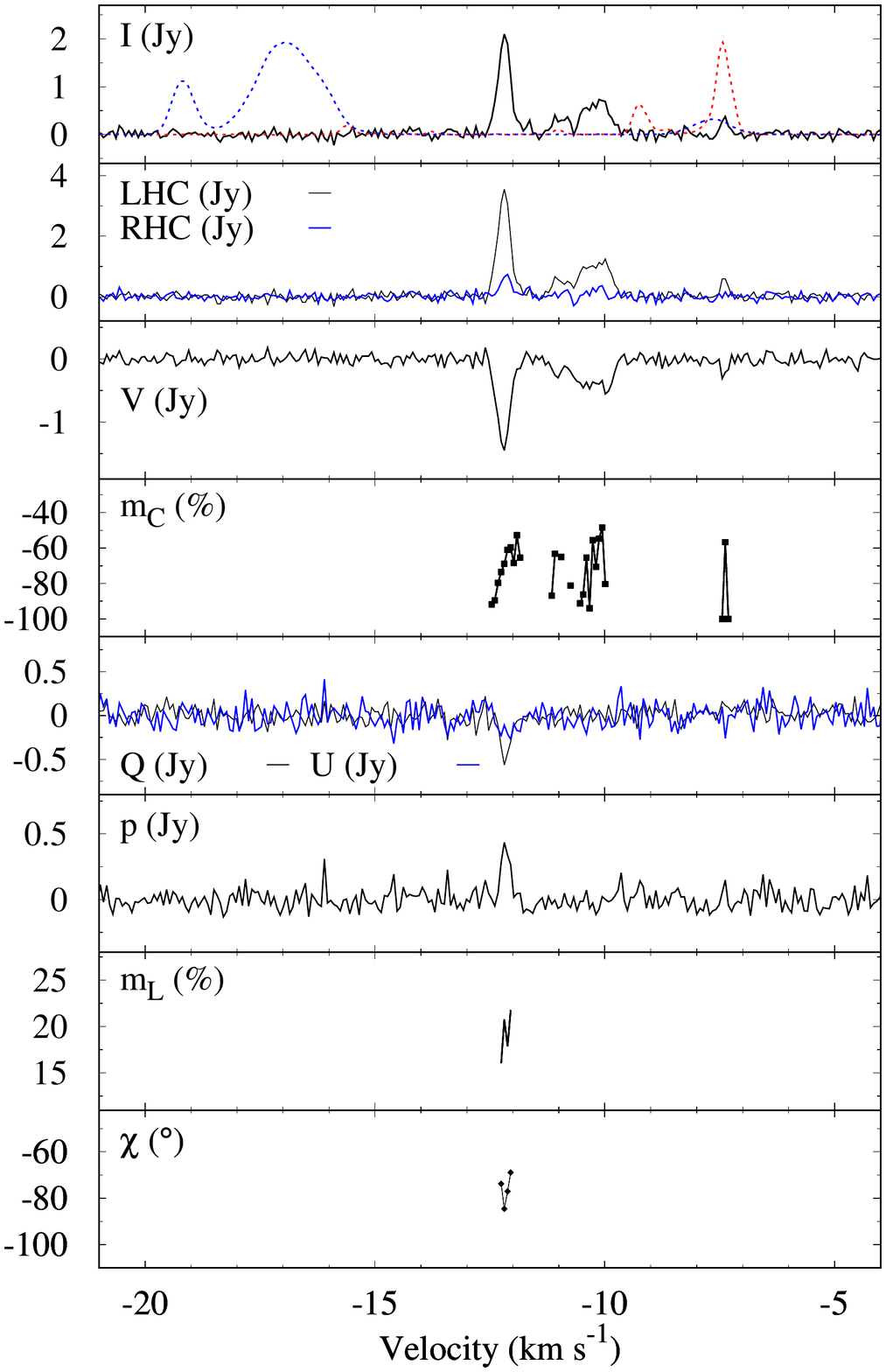}
   \includegraphics[scale=0.5, trim={1.8cm 1.20cm 3.2cm 7.5cm},clip]{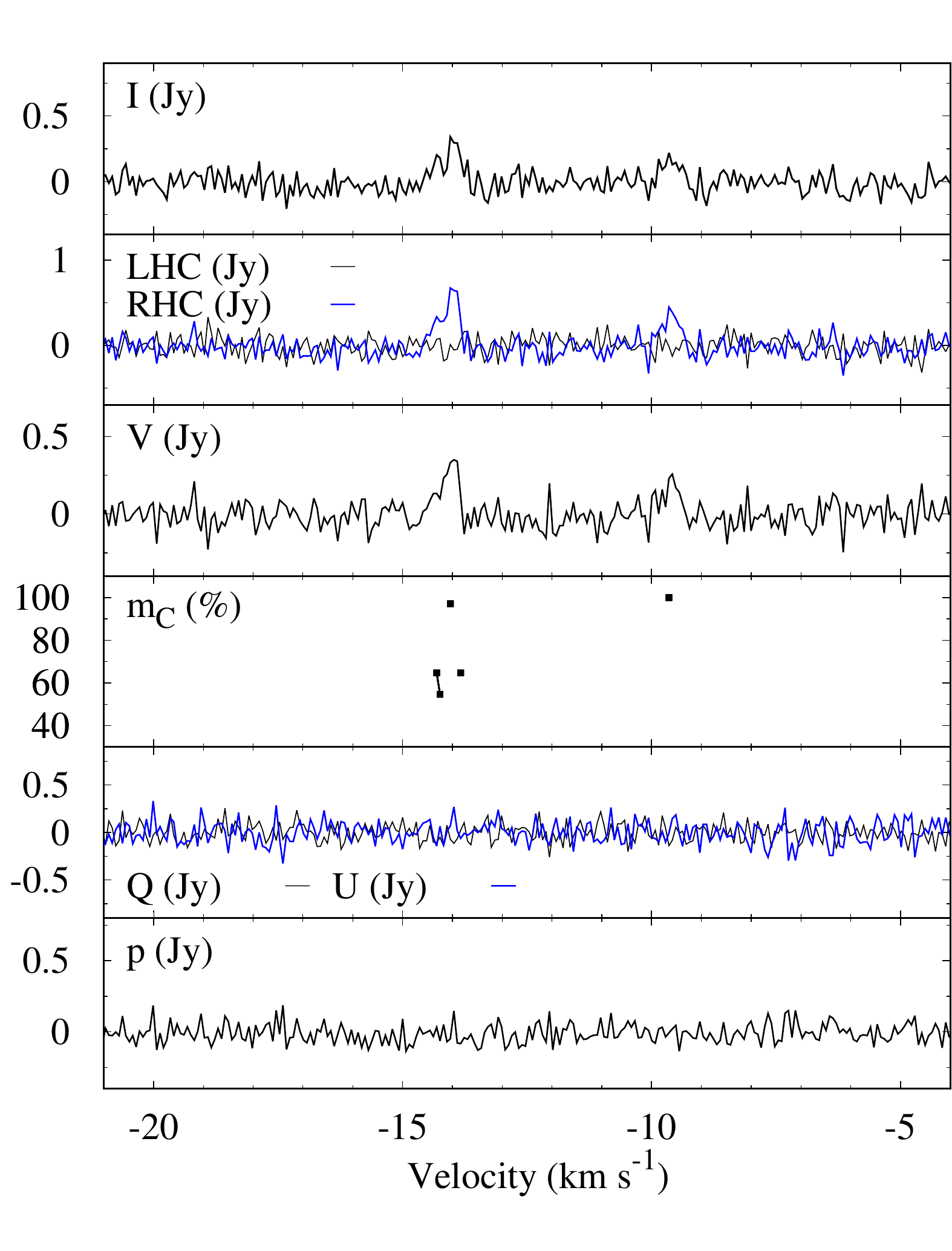}
   \caption{Full polarisation spectra of the OH 1665\,MHz ($left$) and 1667\,MHz ($right$) maser lines of G107 taken on March 9, 2017 (MJD 57821). Spectra of the Stokes $I$, LHC,  RHC, Stokes $V$, degree of circular polarisation ($m_\mathrm{C}$), Stokes $Q$, $U$, linearly polarised flux density ($p$), degree of linear polarisation ($m_\mathrm{L}$), and polarisation position angle $\chi$ are shown from top to bottom.
   In the top left panel the Stokes $I$ spectra of CH$_3$OH (red dotted line) and H$_2$O (blue dotted line) maser lines divided by a factor of 50 and 200, respectively, are added. 
   }
\label{pol-OH1665}
\end{figure}

%

  \begin{figure*}
   \centering
   \includegraphics[width=0.98\textwidth]{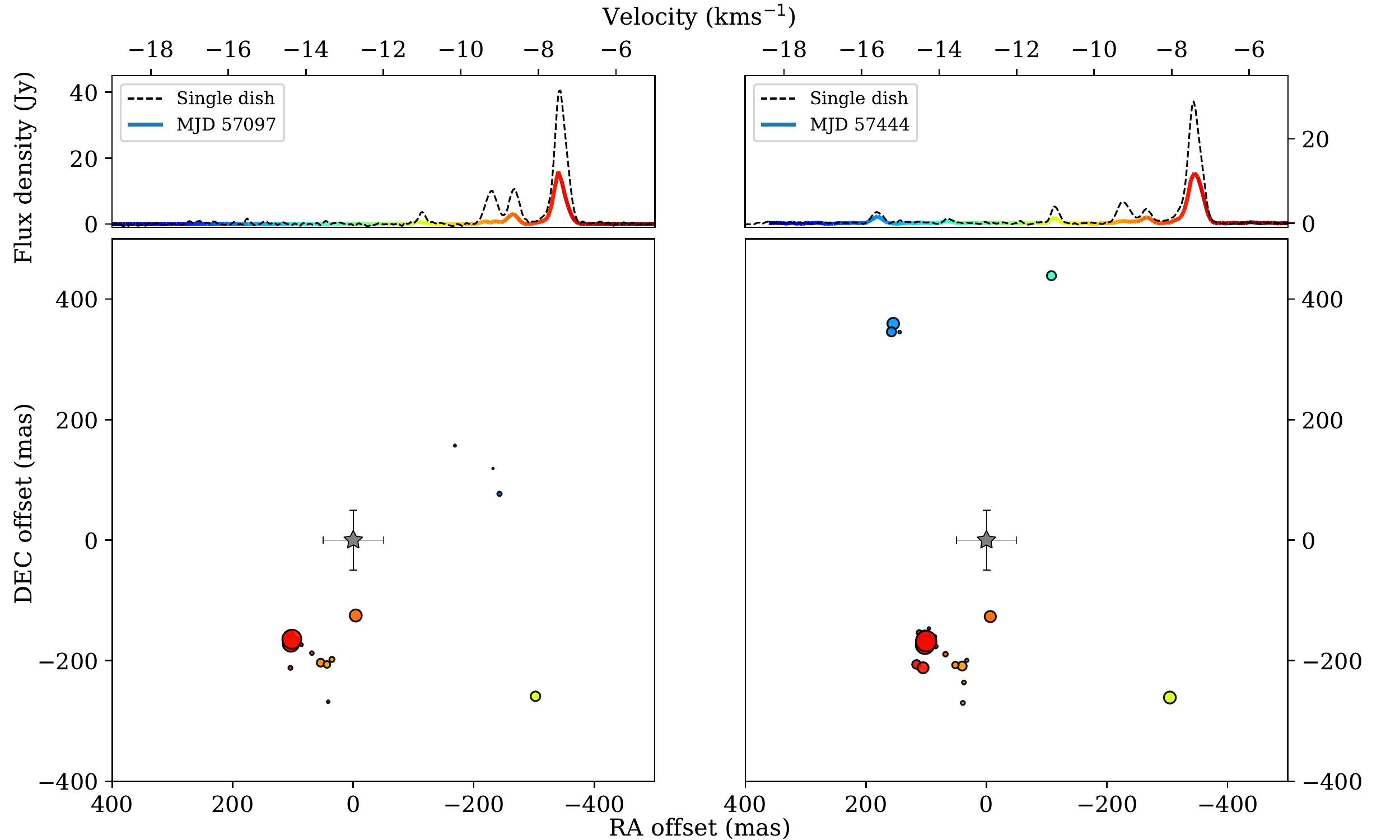}
   \includegraphics[width=0.98\textwidth]{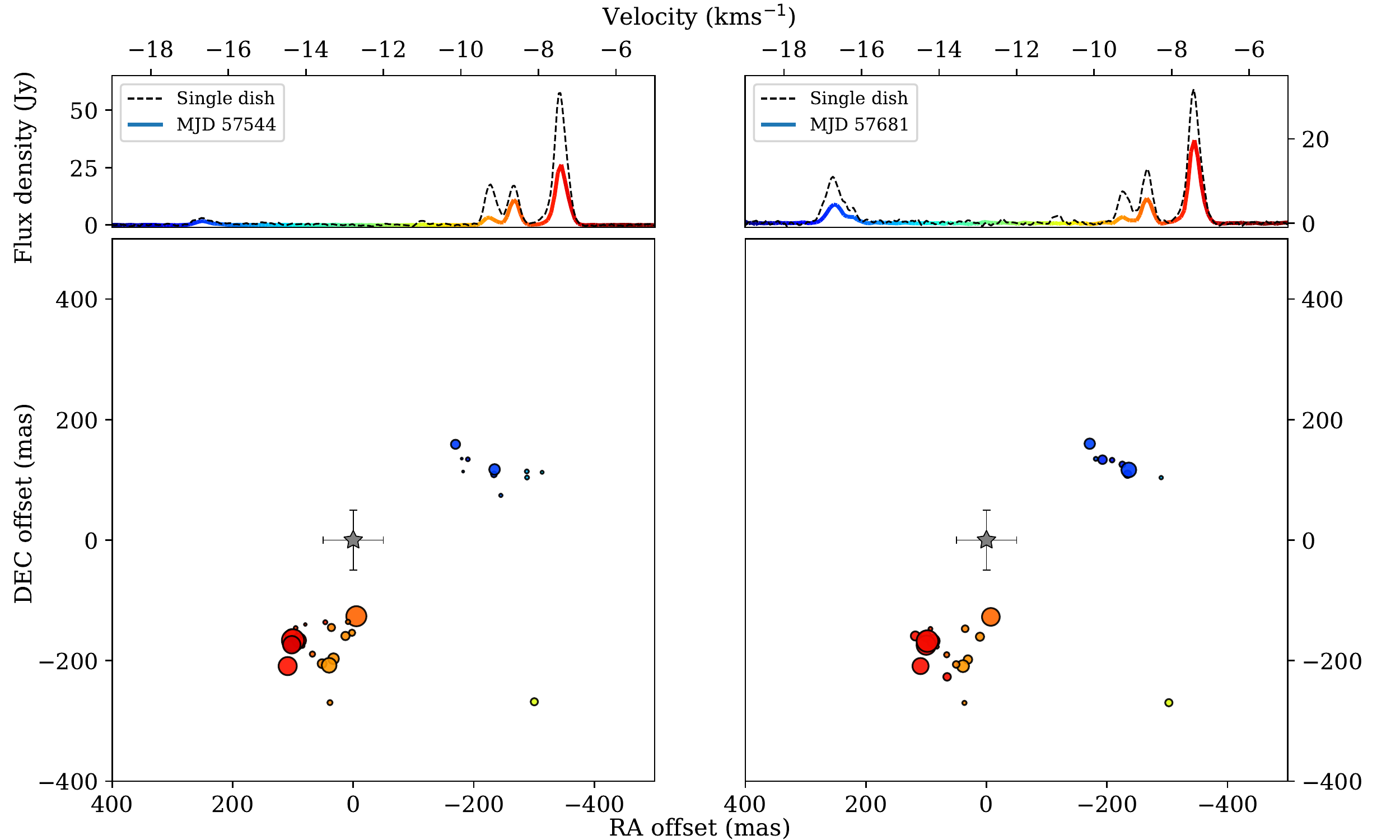}
   \caption{6.7\,GHz maser spectra and spatial structures during all epochs of EVN observations. Coloured lines show the cross-correlated VLBI spectra and colour corresponds to velocity of maser cloudlets shown in the bottom panels. The emission is filtered out by the EVN beam by more than 40-60\% for most of the spectral features. The size of each circle is scaled with the logarithm of peak flux density. The symbol of grey star with the error bars indicates the position of the 1.3\,mm continuum emission peak (\citealt{Palau2013}).
   }
\label{comparison_methanol}
\end{figure*}

\begin{figure*}
   \centering
   \includegraphics[width=0.98\textwidth]{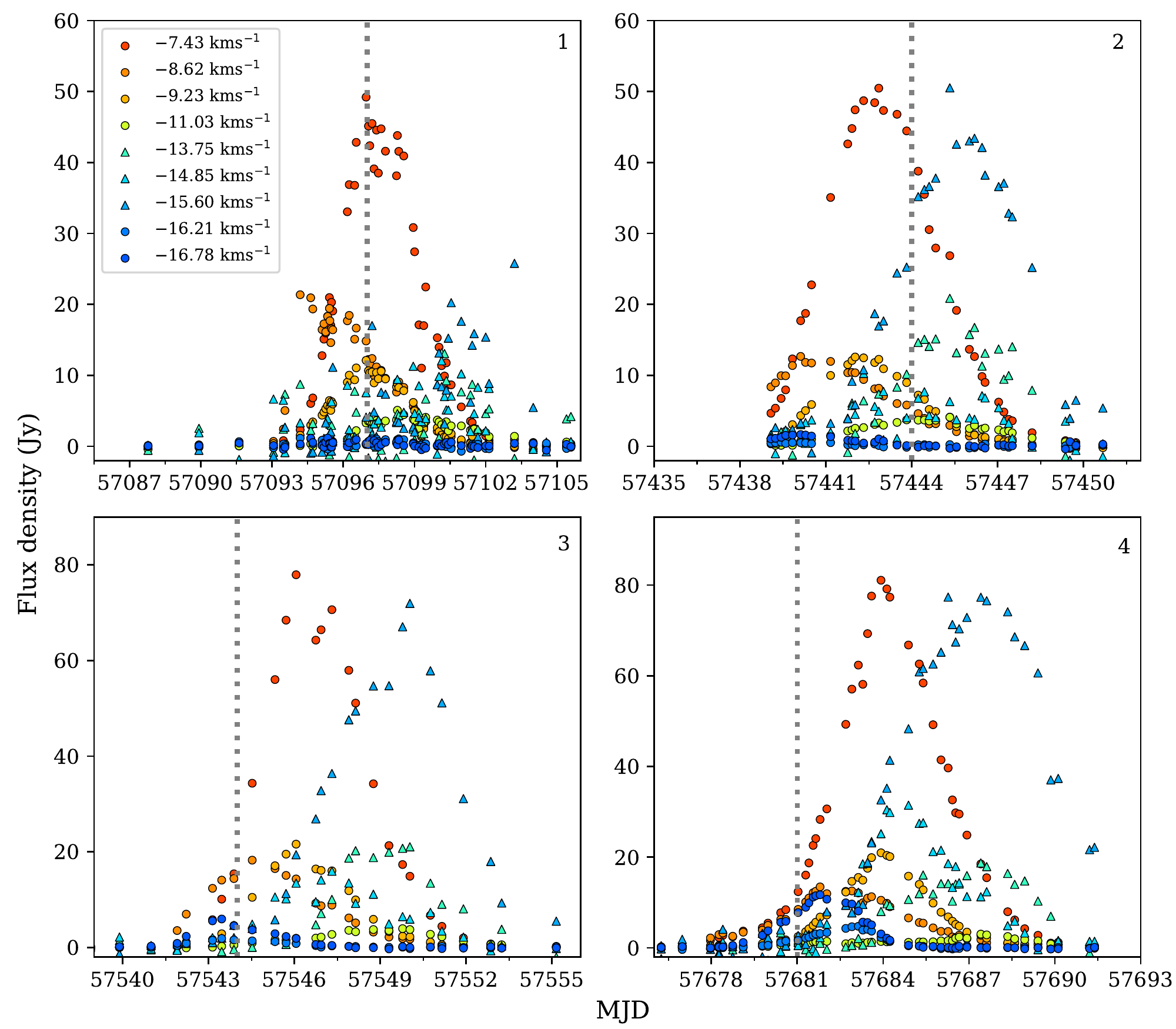}
   \caption{Light curves of 6.7\,GHz maser features around EVN observations. The dotted lines indicates the dates of EVN experiments numbered in the top right corner. The features in which emission was detected only at the second epoch are denoted by triangles and their flux density was scaled up by a factor of 15. Differences in the maser morphology seen by EVN (Fig.~\ref{comparison_methanol}) are clearly related to the flare phase. }
\label{comparison_methanol-2}
\end{figure*}

\begin{figure*}
   \includegraphics[width=0.95\textwidth]{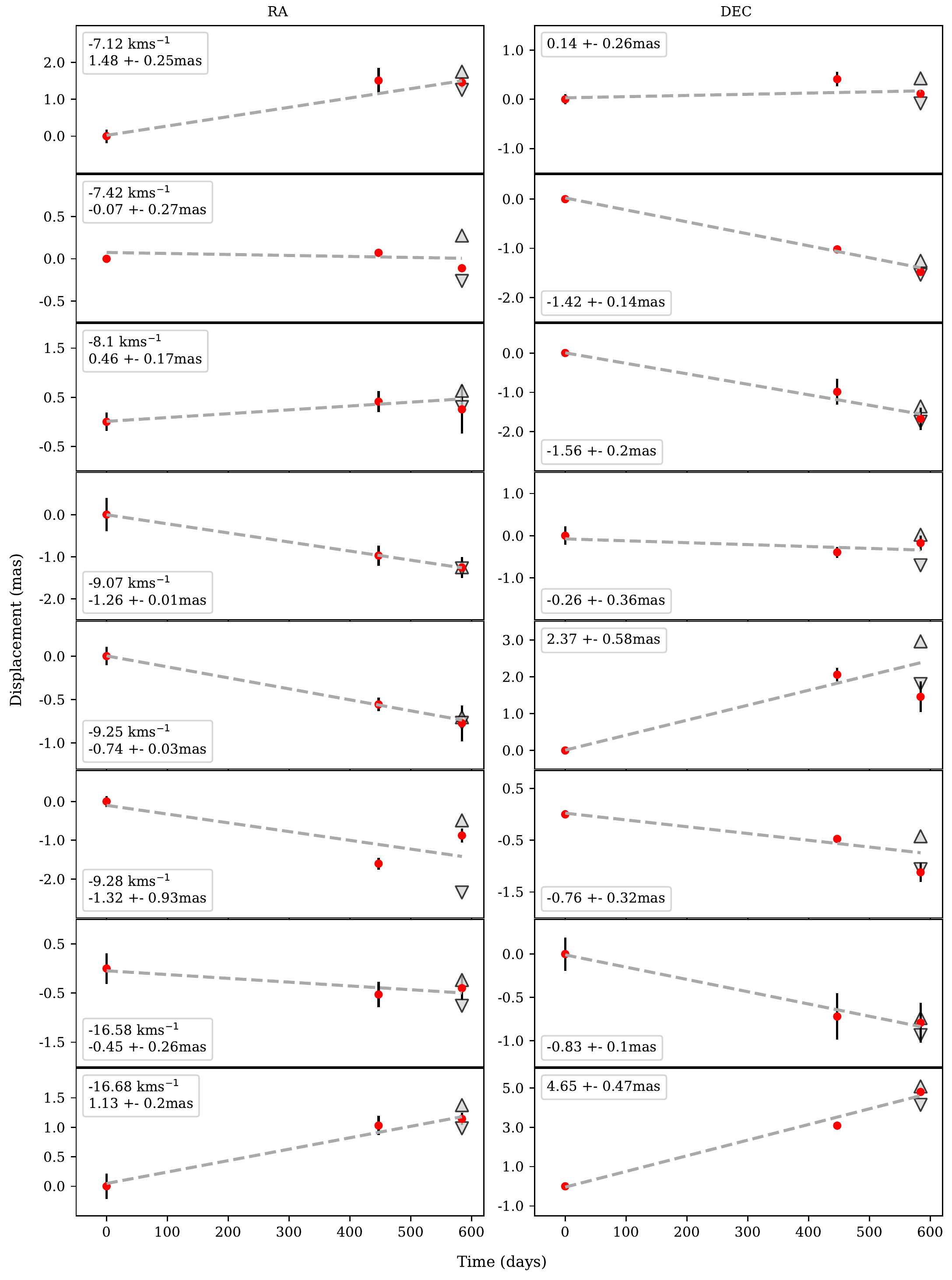}
   \caption{Proper motion measurements vs. time for each cloudlet with RA and DEC components  on the left and right panels, respectively. Red points show measured values with respective errors. Dashed lines show best fits of linear functions. Top and bottom triangles show the error range of the fit. The proper motion with estimated errors for each axis is shown in boxes. The velocity of the component concerned is given in the left panel. The final result is shown on Fig. \ref{fig:prop_mot}.}
\label{fig:prop_jux}
\end{figure*}


\clearpage
\onecolumn

\setcounter{table}{0}
\begin{longtable}{l r r c r c c }
    \caption{Parameters of the cloudlets during the three maser transitions observed over a timespan of 1.1\,yr. Coordinates ($\Delta$RA, $\Delta$DEC) are relative to the map origin in Fig.~\ref{fig:vlbi_comp}. Most of cloudlets are parametrized with the fitted brightness ($S_{\mathrm{f}}$) and line width at half maximum (FWHM). The 6.7\,GHz maser cloudlets seen at epoch 2 are added to show their different appearance relative to that at the three other epochs.\\ \label{tab:vlbi_char}}
    \\
    \hline
    & $\Delta$RA & $\Delta$DEC & $V_\mathrm{p}$ & FWHM & $S_\mathrm{p}$ & $S_\mathrm{f}$ \\
    & (mas) & (mas) & (\kms) & (\kms) & (Jy~beam$^{-1}$)  & (Jy~beam$^{-1}$)\\
    \hline
    \endfirsthead
\caption{Continued.}\\
\hline
  & $\Delta$RA & $\Delta$DEC & $V_\mathrm{p}$ & FWHM & $S_\mathrm{p}$ & $S_\mathrm{f}$ \\
    & (mas) & (mas) & (\kms) & (\kms) & (Jy~beam$^{-1}$)  & (Jy~beam$^{-1}$)\\
\hline
\endhead
\hline
\endfoot
\hline
\endlastfoot
\\
    \multicolumn{7}{c}{\bf OH 1665\,MHz (VLBA)}\\
     & $-$14.954 & $-$294.106 & $-$7.42 &  0.146 &  0.310 & 0.318 \\
    & $-$119.285 & 135.742 & $-$10.27 & 0.892 & 0.525  & 0.484 \\
    & $-$65.271 & 373.965 & $-$12.12 & $-$ & 0.22 & $-$\\
    & 61.145 & 247.472 & $-$12.22 & 0.314 & 1.461 & 1.547 \\
    \multicolumn{7}{c}{\bf OH 1665\,MHz (EVN)}\\
     & $-$9.774 & $-$300.027 & $-$7.56 & $-$ & 0.096 & $-$ \\
    & $-$115.510 & 130.847 & $-$10.40 & 1.130 & 0.485 & 0.479 \\
    & $-$49.913 & 372.111 & $-$12.00 & $-$ & 0.050 & $-$\\
    & 67.252 & 241.243 & $-$12.33 & 0.567 & 0.744 & 0.773 \\
    \\
    \multicolumn{7}{c}{\bf H$_2$O 22.2\,GHz} \\
     & $-$13.313 & $-$74.900 & $-$7.58 & 1.186 & 21.543 & 12.968 \\
    & $-$204.755 & 41.600 & $-$15.97 & 1.048 & 29.513 & 30.837 \\
    & $-$202.988 & 36.955 & $-$16.20 & 0.813 & 32.567 & 32.353  \\
    & $-$205.331 & 40.530 & $-$16.60 & 0.804 & 82.596 & 82.593 \\
    & $-$197.815 & 39.607 & $-$16.71 & $-$ & 3.789 & $-$\\
    & $-$181.082 & 7.593 & $-$16.90 & 1.331 & 18.218 & 6.635 \\
    & $-$205.179 & 39.908 & $-$17.30 & $-$ & 27.191 & $-$ \\
    & $-$195.511 & 30.094 & $-$17.31 & 0.705 & 41.828 & 40.240 \\
    & $-$179.586 & 5.135 & $-$17.44 & 0.790 & 42.912 & 40.588 \\
    & $-$179.410 & 3.670 & $-$19.17 & 0.559 & 40.56 & 40.385 \\
\\
    \multicolumn{7}{c}{\bf CH$_3$OH 6.7\,GHz epoch 4 (MJD 57681)}\\
    & 99.886 & $-$174.277 & $-$7.13 & $-$ & 8.865 & $-$ \\
    & 88.092 & $-$167.362 & $-$7.21 & 0.287 & 1.741 & 1.656 \\
    & 97.987 & $-$167.545 & $-$7.36 & 0.332 & 15.630 & 16.429 \\
    & 117.979 & $-$159.029 & $-$7.39 & 0.246 & 0.652 & 0.674 \\
    & 65.320 & $-$226.854 & $-$7.40 & $-$ & 0.434 & $-$ \\
    & 109.386 & $-$209.002 & $-$7.40 & 0.275 & 3.782 & 3.891 \\ 
    & 100.701 & $-$159.262 & $-$7.73 & 0.295 & 0.969 & 0.939\\
    & 92.954 & $-$147.215 & $-$7.85 & $-$ & 0.105 & $-$\\
    & 82.251 & $-$176.911 & $-$8.06 & 0.442 & 0.101 & 0.101 \\
    & 65.986 & $-$190.234 & $-$8.55 & 0.311 & 0.189 & 0.200 \\
    & $-$7.284 & $-$127.389 & $-$8.59 & 0.280 & 5.807 & 5.829 \\
    & 35.387 & $-$147.094 & $-$8.99 & 0.287 & 0.338 & 0.329 \\    
    & 50.225 & $-$206.324 & $-$9.01 & 0.292 & 0.310 & 0.320\\
    & 36.559 & $-$270.120 & $-$9.08 & $-$ & 0.129 & $-$ \\
    & 10.817 & $-$160.301 & $-$9.12 & 0.122 & 0.515 & 0.517 \\
    & 30.766 & $-$198.142 & $-$9.20 & 0.211 & 0.569 & 0.568 \\
    & 38.960 & $-$209.079 & $-$9.26 & 0.228 & 1.389 & 1.420\\
    & $-$302.581 & $-$269.756 & $-$10.91 & 0.215 & 0.328 & 0.382 \\
    & $-$289.855 & 103.608 & $-$14.95 &$-$ & 0.082 & $-$ \\
    & $-$181.412 & 134.757 & $-$16.33 & 0.598 & 0.112 & 0.114 \\
    & $-$171.342 & 159.929 & $-$16.47 & 0.749 & 1.102 & 0.956 \\
    & $-$236.113 & 116.447 & $-$16.58 & 0.510 & 2.974 & 2.859 \\
    & $-$234.022 & 109.275 & $-$16.87 & $-$ & 0.412 & $-$\\
    & $-$225.448 & 125.518 & $-$16.90 & 0.364 & 0.230 & 0.262 \\    
    & $-$192.541 & 133.441 & $-$16.91 & 0.285 & 0.595 & 0.592 \\
    & $-$208.412 & 132.691 & $-$17.11 & 0.361 & 0.165 & 0.154 \\
 \\   
     \multicolumn{7}{c}{\bf CH$_3$OH 6.7\,GHz epoch 2 (MJD 57444)}\\
    & 89.126 & $-$168.899 & $-$7.05 & 0.298 & 0.370 & 0.384 \\
    & 102.004 & $-$173.640 & $-$7.12 & 0.294 & 7.132 & 7.243 \\
    & 111.537 & $-$153.678 & $-$7.16 & 0.252 & 0.197 & 0.206 \\
    & 116.095 & $-$206.232 & $-$7.31 & 0.311 & 0.599 & 0.629 \\
    & 100.052 & $-$167.717 & $-$7.35 & 0.367 & 11.853 & 12.159 \\
    & 105.166 & $-$211.788 & $-$7.50 & 0.220 & 1.094 & 1.113 \\
    & 102.335 & $-$159.127 & $-$7.79 & $-$ & 1.170 & $-$ \\ 
    & 95.592 & $-$146.682 & $-$7.85 & $-$ & 0.052 & $-$ \\
    & 84.405 & $-$176.262 & $-$8.04 & 0.337 & 0.114 & 0.117 \\
    & 95.083 & $-$162.762 & $-$8.19 & $-$ & 0.057 & $-$ \\
    & 85.779 & $-$159.710 & $-$8.50 & 0.343 & 0.058 & 0.058 \\
    & 67.988 & $-$189.282 & $-$8.54 & 0.262 & 0.155 & 0.156 \\
    & $-$6.336 & $-$126.889 & $-$8.59 & 0.293 & 1.195 & 1.197 \\
    & 27.436 & $-$236.224 & $-$8.92 & $-$ & 0.107 & $-$ \\
    & 51.579 & $-$207.241 & $-$9.01 & 0.202 & 0.313 & 0.309 \\
    & 39.190 & $-$270.110 & $-$9.08 & $-$ & 0.115 & $-$ \\
    & 32.588 & $-$199.424 & $-$9.25 & $-$ & 0.077 & $-$ \\
    & 40.285 & $-$208.908 & $-$9.28 & 0.223 & 0.669 & 0.650 \\
    & 304.259 & $-$261.180 & $-$10.96 & 0.227 & 1.408 & 1.427 \\
    & $-$107.812 & 438.657 & $-$13.70 & 0.223 & 0.645 & 0.670 \\
    & 144.067 & 345.034 & $-$14.81 & $-$ & 0.054 & $-$ \\
    & 154.643 & 359.164 & $-$15.50 & 0.267 & 1.269 & 1.330 \\
    & 157.612 & 345.580 & $-$15.63 & 0.238 & 0.707 & 0.709\\   
    \hline
\end{longtable}

\end{appendix}

\end{document}